\documentclass[apj]{emulateapj}
\usepackage{natbib}
\usepackage{amsmath}
\usepackage{graphicx}
\usepackage{txfonts}
\usepackage[scaled]{helvet}
\usepackage{epsfig}
\usepackage[normalem]{ulem}
\usepackage{color}

\begin{document}

\title{The effect of spots on the luminosity spread of the Pleiades}

\author{Zhen Guo$^{1,\, 2,\,3}$, Michael Gully-Santiago$^4$, Gregory J. Herczeg$^1$}

\altaffiltext{1}{Kavli Institute for Astronomy and Astrophysics,
  Peking University, Yi He Yuan Lu 5, Haidian District, Beijing 100871,
  P.R. China; z.guo4@herts.ac.uk}
\altaffiltext{2}{Department of Astronomy, School of Physics, Peking
  University, Yi He Yuan Lu 5, Haidian District, Beijing 100871,
  P.R. China}
  \altaffiltext{3}{Centre for Astrophysics Research, University of Hertfordshire, Hatfield AL10 9AB, UK}
\altaffiltext{4}{NASA Ames Research Center, Moffett Field, CA 94035, USA}

\begin{abstract}
Cool spots on the surface of magnetically-active stars modulate their observed brightnesses and temperatures, thereby affecting the stellar locus on the H-R diagram. Recent high precision space-based photometric surveys reveal the rotational modulation from spots on stars in young clusters, including K2 monitoring of the 125-Myr-old Pleiades cluster. However, lightcurves reveal only the asymmetries in the visible spot distributions rather than the total sizes of spots on stellar surfaces, which causes a discrepancy between the spot coverage measured by photometric and spectroscopic observations. In this paper, we simulate photometric variability introduced by randomly-distributed starspots on a 125-Myr-old coeval cluster.  Our simulation results show that randomly distributed small spots on the stellar surface would explain this discrepancy that the photometric observations only reveal 10\% to 40\% of the spot coverage measured by spectra. The colors and luminosities of photospheres are modeled for a range of photospheric temperature, spot coverage, and spot temperature.  The colors and luminosities of a simulated population are then compared to the luminosity spread of Pleiades members, excluding the 25\% of candidates that are identified as non-members with Gaia DR2 astrometry. The observed luminosities of Pleiades members have a standard deviation of 0.05 dex, which could be entirely explained by spots with a star-to-star standard deviation of spot coverage of 10\%, but with an average coverage area that is not well constrained.  

\end{abstract}
\shortauthors{Guo et al.}
\accepted{October 23, 2018}

\section{Introduction}

Starspots are the visible manifestation of internal magnetic activity \citep[see reviews by][]{Schrijver2000,Strassmeier2009}. The magnetic dynamos on the sun and other stars are produced by the rotation of convective plasma \citep{Browning2010, Charbonneau2014}. Single stars rotate fastest when they are young, generating strong magnetic activity \citep[see review by][]{Bouvier2014}. While solar magnetic activity has a negligible effect on the total irradiance \citep{Balmaceda2009} and flux transport of the sun, at young ages the internal magnetic fields can change the structure of the star and the efficiency of convection \citep{Somers2015, Feiden2016, MacDonald2017}.  One of the consequences of this magnetic activity, starspots, also complicates the measurement of stellar properties by introducing additional temperature components.

Since magnetically-active stars are rarely resolved \citep[see, e.g.,][]{Roettenbacher2016}, starspots are usually detected with photometric monitoring \citep[e.g.][]{Hall1972, Strassmeier1997, Meibom2009}.  The recent space-based COROT and  Kepler missions provide sensitive lightcurves to hunt for exoplanets \citep{Baglin2003,Borucki2010}, and can also be used to measure stellar rotation periods for analyzing angular momentum evolution \citep[e.g.][]{Matt2015}. The rotation amplitudes of the lightcurves are measured as consequences of star spots \citep{Savanov2017}. In addition, the stellar surfaces are mapped by lightcurve inversion techniques that infer the spot geometry and evolution through the morphologies of lightcurves \citep[e.g.][]{Savanov2012, Roettenbacher2013}. The K2 lightcurves of stars in the  125-Myr-old Pleiades cluster typically have amplitudes of 1--10\% \citep{Rebull2016a}, less than the 30--50\% amplitudes often seen in at least some 1--10 Myr stars \citep[e.g.][]{Grankin2008,Alencar2010,Lanza2016}. However, the total spot coverage only from lightcurve amplitudes may be underestimated because symmetric morphologies, such as polar spots, will not cause variability \citep[see discussions in][]{Rebull2016b,Rackham2018}.

In contrast to lightcurves, spectroscopic techniques can provide estimates of the total spot coverage by comparing (usually molecular) features of a star with those from an unspotted template \citep[e.g.][]{Petrov1994, Neff1995, ONeal1996, Fang2016, Gully2017}.  The magnetic fields themselves can be measured either through Zeeman broadening, which yields an averaged magnetic field strength over the visible stellar surface \citep[e.g.][]{Johnskrull2004,Lavail2017}, or using polarized light, which when combined with high-resolution spectroscopy  yields maps of the largest magnetic structures \citep[Zeeman Doppler Imaging; see review by][]{Donati2009}.

These detection methods all leverage the effect that spots have on the emission from the star.  While many studies have investigated the subsequent limitations on our ability to measure the presence of planets \citep[e.g.][]{Desort2007, Reiners2010} and exoplanet atmospheres \citep{Rackham2018}, the spots also interfere with our ability to measure properties of spotted young stars.  Spots alter the evolution of pre-main sequence stars \citep{Somers2015}, require radius inflation relative to predictions from standard models to emit the same amount of energy \citep[e.g.][]{Jackson2013,  Somers2017}, and introduce cooler components into the emission from the star \citep{Gully2017}.  Differences in spot properties among stars may introduce a spread in the observed properties of co-eval cluster members, which could help to explain the luminosity spreads seen to every star-forming region \citep[see reviews by][]{Preibisch2012, Soderblom2014}.

In this work, we investigate spot properties of stars in the Pleiades cluster ({\it Subaru} in Japanese and {\it M{\v a}o} in Chinese), the closest young open cluster with a low foreground extinction \citep{Mermilliod1981}.  The Pleiades has an age of 125 Myr, as estimated by the lithium depletion boundary \citep[e.g.][]{Stauffer1998}, which places this cluster near the peak of the angular velocity evolution of low-mass stars \citep[see review by][]{Bouvier2014}.  In Section 2, we describe differences between photometric and spectroscopic measurements of spots on the Pleiades.  In Section 3, we simulate lightcurves for different spot properties and demonstrate that the spectroscopic and photometric differences may be explained if these stars have many small spots.  In Section 4, we then investigate how spots affect the loci of low-mass young stars in color-magnitude diagrams and discuss observational biases introduced when starspots are not considered.  In Section 5, we explain the observed luminosity spread in the Pleiades cluster by the appearance of cool spots. We then summarize our results in Section 6.

\begin{figure}[!t]
\includegraphics[width=3.5in,angle=0]{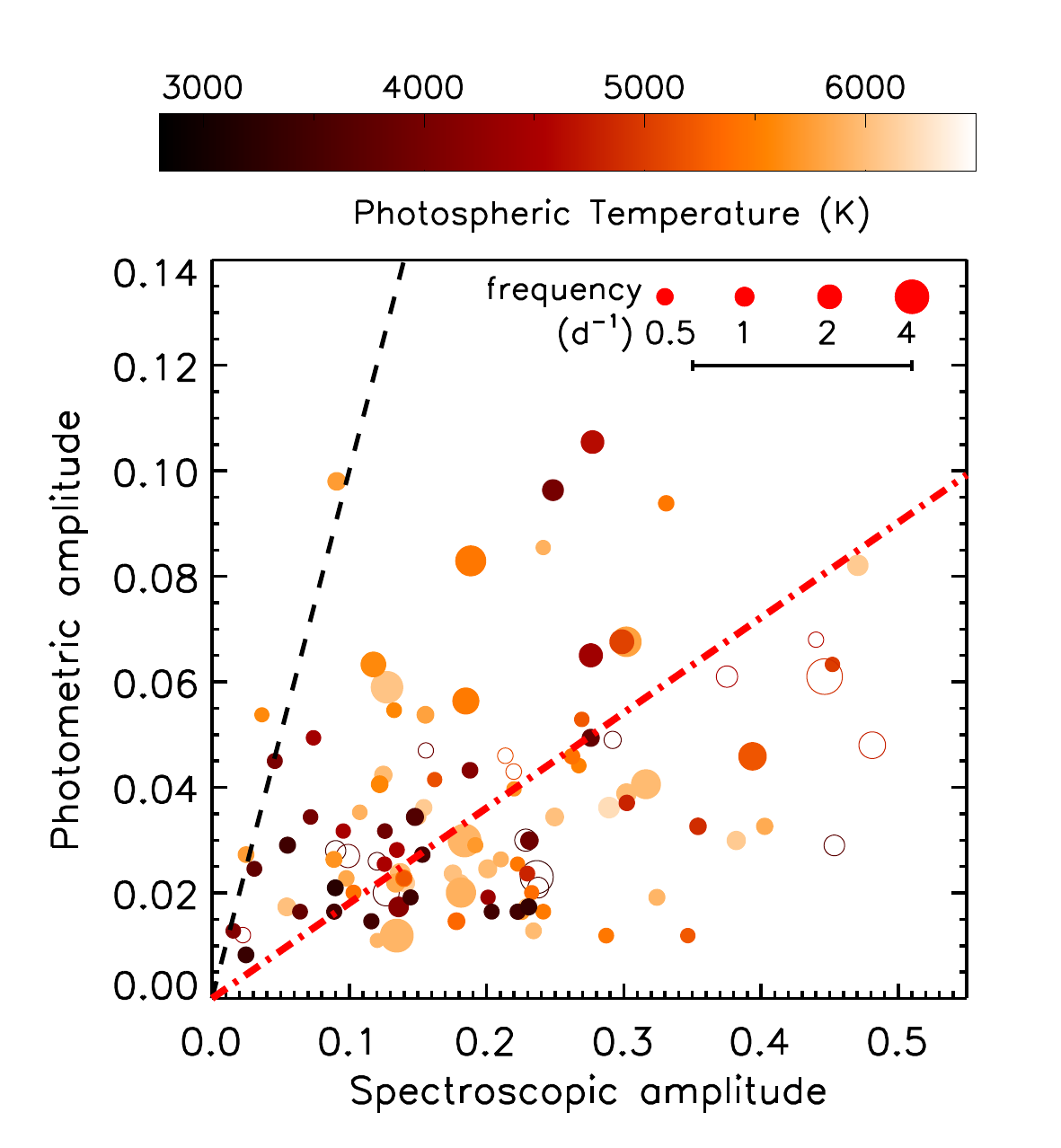}
\caption{Spot coverage of Pleiades members inferred from photometry and spectroscopy. The photometric amplitude is directly adopted from 10 - 90 percentile amplitude of K2 lightcurves from \citet{Rebull2016a}. The spectroscopic amplitude is calculated based on the $f_{\rm spot}$, $T_{\rm spot}$ and $T_{\rm eff}$ from  LAMOST spectra \citep{Fang2016}. The color of each point represents the photospheric temperature derived by color-magnitude diagram from \citet{Fang2016}, and the sizes of the symbols are proportional to the rotation frequency of K2 lightcurves (0.07 to 4.25 $d^{-1}$, see examples at the upper right corner). The membership classification is directly adopted from \citet{Kamai2014}. The Pleiades members are shown by filled dots, while non-member and binaries are shown by open circles. The black dashed line represents the photometric amplitude equal to spectroscopic amplitude, and the red dot-dashed line is the median ratio between the two amplitudes. The error bar shown in the upper right corner is a standard 10\% error in spot filling factor from \citet{Fang2016}. }
\label{fig:k2_lamost}
\end{figure}

\section{Observations of spots on stars in the Pleiades}

As the nearest young open cluster, the Pleiades is a cornerstone for studies of the early evolution of young stars and their rotation.  The cluster includes $\sim 2100$ known members \citep{Stauffer2007, Lodieu2012, Bouy2015} with a mass function described by a log-normal distribution and a mean characteristic mass of 0.2 M$_\odot$ \citep{Hambly1999, Moraux2003, Moraux2004, Deacon2004,Lodieu2012}.  The distances of the Pleiades members have an average of $137$ pc from Gaia DR2 parallaxes with a $4$ pc standard deviation that includes measurement uncertainties and a real depth \citep{Gaia2018dr2catalog}. Spots on Pleiades stars have been extensively detected through photometric monitoring \citep[e.g.][]{vanLeeuwen1987, Stauffer1987b, Hartmann2010, Covey2016}.  Redder $V - K_s$ colors\footnote{ In this paper, we apply the optical bands ($B, V, I$) from Johnson photometric system and infrared $K_s$-band from 2MASS system} are observed on fast rotating low mass members, suggesting cool spots or stronger magnetic activities on fast rotators. In this section, we first revisit the photometric and spectroscopic observations for the Pleiades members, then compare these results to obtain a geometric view of the starspots on the Pleiades members.   

The photometric lightcurves for the Pleiades members analyzed here were obtained in the K2 extension \citep{Howell2014} of the Kepler Space Telescope. The 72 days long K2 monitoring campaign (C04, Feb 07--Apr 23, 2015) included 826 candidate Pleiades members, of which 92\% have accurate periods measured from spot modulation \citep{Rebull2016a}\footnote{\citet{Rebull2016a} suggest that the 8\% of members that lack periodicity have lightcurves that are likely affected by non-astrophysical contributions \citep{Vanderburg2014}.}. For low mass stars, the photometric amplitude, $\Delta F_{\rm phot}$, defined as the 10 to 90 percentile range of the K2 lightcurve, is typically 1--10\% of the average emission. Estimating the spot coverage from photometric lightcurves alone captures solely asymmetric structures, resulting in significant underestimates of starspot area \citep{Rackham2018}. {  
In this work, we adopt the photometric amplitudes of Pleiades members measured by \citet{Rebull2016a} from K2 lightcurves obtained in Cycle 4.
}

{  The spot contributions for 304 Pleiades members were measured by \citet{Fang2016} through low-resolution  ($R \sim 1000$) spectra spanning 3700--9000 \AA\ that were obtained with LAMOST \citep[Large sky Area Multi-Object fiber Spectroscopic Telescope;][]{Zhao2012}. The spot temperature and covering area were measured based on empirical relationships between TiO absorption bands ratios. \citet{Fang2016} fit the observed TiO features by a warmer photospheric temperature, derived from $V - I$ colors, and a cooler spot temperature left as a free parameter. This spot measurement includes a degeneracy between spot coverage and spot temperature, so \citet{Fang2016} adopt the minimum spot size that can explain the TiO feature depths, hence a lower limit of spot temperature. The LAMOST-based estimates of $f_{\rm spot}$ are more uncertain when the stellar effective temperature ($T_{\rm eff}$) is smaller; for example \citet{Fang2016} reports typical measurements of $0-50\%$, with comparable uncertainties of 10\% for $T_{\rm eff} > 3800$ K and  $> 15\%$ for $T_{\rm eff} < 3800$ K. The measured spot parameters from \citet{Fang2016} is provided through private communication. The photometric observations of spots in the Pleiades are combined in this paper.} The spectroscopic amplitude ($\Delta F_{\rm spec}$) is defined here, based on the spot and stellar parameters from the  LAMOST spectra \citep{Fang2016}, as
\begin{equation}
\Delta F_{\rm spec} =  f_{\rm spot}(1- F_{\rm spot}/F_{\rm phot})
\end{equation}
where $f_{\rm spot}$ represents the spot coverage. $F_{\rm spot}$ and $F_{\rm phot}$ are the fluxes from the spot and photospheric regions in a unit area through the {\it Kepler}-band, both are generated from BT-Settl models with solar metallicity (see \S3.1 for more information). The spectroscopic amplitude $\Delta F_{\rm spec}$ corresponds to the change in the absolute brightness introduced by the spots at the time of the observation. The definition of $\Delta F_{\rm spec}$ is slightly mismatched $\Delta F_{\rm phot}$. In the extreme scenario where the photometric amplitude is described by a single spot on the visible hemisphere, then $\Delta F_{\rm spec}$ would equal $1.25 \times \Delta F_{\rm phot}$ for a given spot filling factor and temperature.

The ratio between $\Delta F_{\rm phot}$ and $\Delta F_{\rm spec}$ offers a simplified geometrical assessment of the level of symmetry of the spots. 
When the star is covered by spots with a high degree of longitudinal symmetry, $\Delta F_{\rm phot}$ would be much smaller than $\Delta F_{\rm spec}$. For instance, polar spots, circumpolar spots on inclined stars, Jupiter-like bands, and a uniform distribution of small spots would all induce negligible temporal photometric modulation despite the coverage of the stellar surface. On the other hand, in some cases, the spectral amplitude is equal to the photometric amplitude, which indicates that the star may have a single large spot visible on one hemisphere.  Figure~\ref{fig:k2_lamost} compares the spot modulation amplitudes of 113 Pleiades sources that have both a photometric amplitude measured from K2 \citep{Rebull2016a,Rebull2016b} and spot parameters estimated from  LAMOST spectra \citep{Fang2016}.  Most stars sit closer to the symmetric or ``circumpolar spot'' regime than to the ``single equatorial spot'' regime.  The large uncertainty in  LAMOST-based $f_{\rm spot}$ estimation results in an $\pm 8\%$ error bar on $\Delta F_{\rm spec}$ in Figure \ref{fig:k2_lamost}, especially since \citet{Fang2016} used a spot temperature that leads to the minimum spot coverage.  However, when taken in aggregate, the spectroscopic amplitudes are much larger than the photometric amplitudes.  

\begin{figure}[!t]
\centering
\includegraphics[width=3.1in,angle=0]{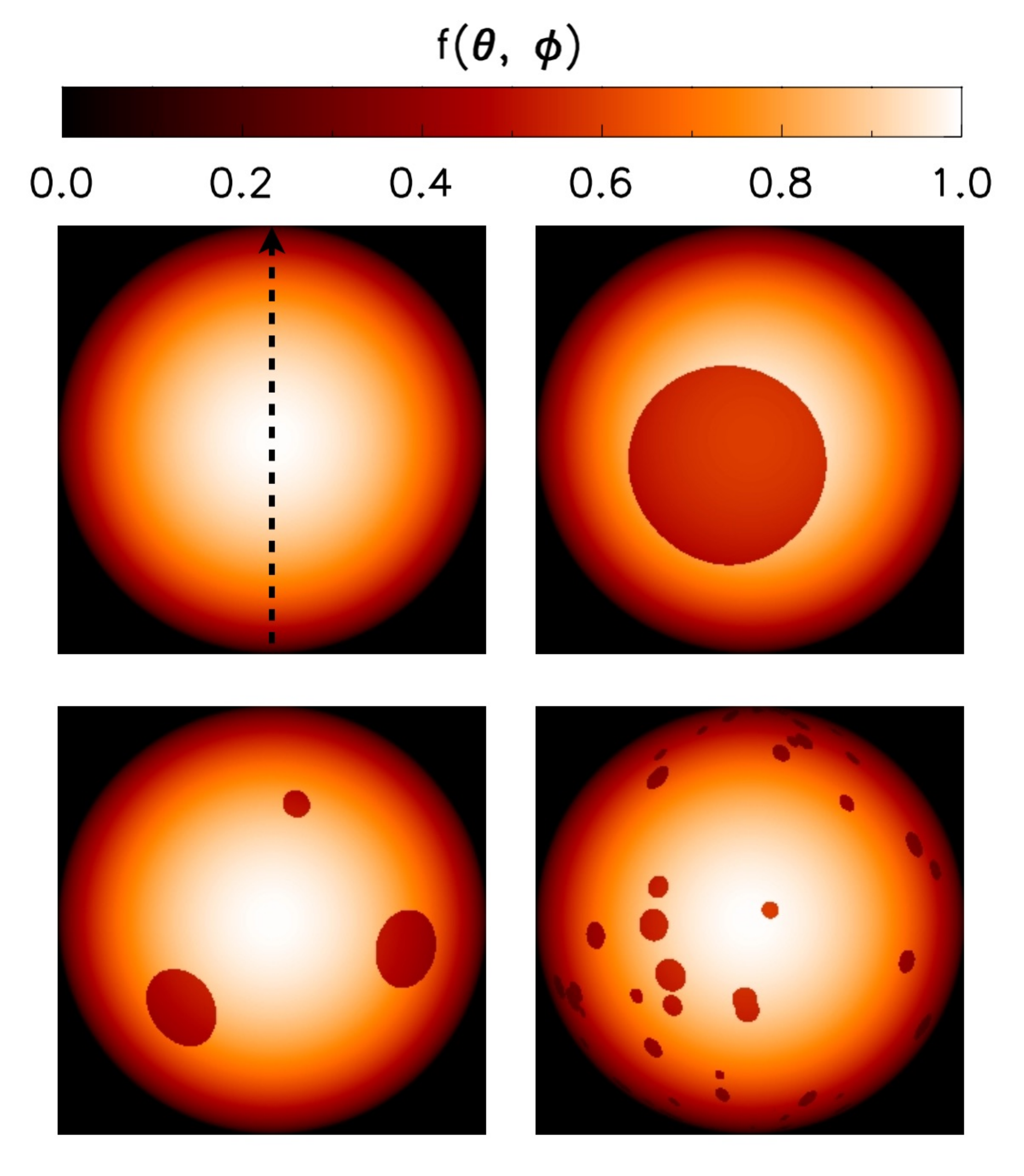}
\caption{Four 2-D maps of the Monte-Carlo simulations, with parameters: $T_{\rm phot} = 4500$ K, $T_{\rm spot} = 4200$ K, and $f_{\rm spot} = 10\%$. {\it Upper left:} a photospheric map with an arrow showing the rotation axis. {\it Upper right:} A map for the ``single spot'' case. {\it Bottom left:} A map for ``multiple spots'', with example of 8 spots on stellar surface, of which 3/8 are located at the front side of the star.  {\it Bottom right:} An example of the ``small spots'' morphology. {  The shapes of spots in the ``single spot'' and ``small spots'' morphologies are set as circular. Spots in the ``multiple spots'' case are ellipse with $e < 0.5$.}}
\label{fig:example_morphology}
\end{figure}

\section{Simulating Spotted Photospheres}

In the previous section, we compared the results from \citet{Fang2016} and \citet{Rebull2016a} to establish that spots cover a large fraction of the surface of young stars yet produce lightcurves with small amplitudes.  In this section, we simulate broadband lightcurves of spotted young stars for a variety of spot distributions and parameters to help interpret this difference between the photometric and spectroscopic measurements of spots.

{  The simulations of inhomogeneous stellar photospheres in this work follow similar methodologies of stellar spots by \citet{Desort2007} and \citet{Rackham2018} to understand their effects on exoplanet detection and characterization. Our Monte-Carlo simulations first generate co-eval 125-Myr-old stellar clusters each contains 2000 stars with masses evenly span from 0.08 to 1.32 solar mass. 
with stellar effective temperatures given by stellar evolutionary models \citep{Baraffe2015}. Then, cool spots with certain temperatures are added on each stellar surface occupying a range of regions and geometries (more details of spot configurations are introduced in Section 3.2).  Finally,  lightcurves in photometric bands ($B, V, R, I, K_s$) are integrated from the synthetic BT-Settl models \citep[CIFIST2011 2015][]{Allard2014} and stellar rotations. }  

We use these simulations to investigate the impact of a spread in spot coverage on the luminosity and color spreads on coeval star clusters. Then, the spreads in spot areal surface coverage are ultimately constrained based on the observed spread in the pre-main-sequence H-R diagram. Hence, the observed luminosity and color spreads may be introduced by existing spots rather than by any bona fide age spread.  In this context, our star spot simulations inform the analysis of H-R diagrams by helping to reveal the different possible configurations that could simultaneously explain the photometric and spectroscopic detection of spots.

\subsection{Stellar samples}

The Pleiades cluster is simulated with 2000 stars at an age of 125 Myr, with temperatures and luminosities adopted from the \citep{Baraffe2015} evolutionary tracks. The stellar masses are distributed between 0.08--1.32 $\rm M_\odot$\footnote{The  LAMOST survey of the Pleiades cluster restricted to $M_\star >$ 0.3 $\rm M_\odot$} to span the range from the approximate detection limit of K2 to the approximate temperature above which spot modulation mixes with stellar pulsations \citep{Balona2015}. {  The inclination angles ($i, j$) of the stellar axes are randomly assigned between $0^\circ < i < 180^\circ$ and $-90^\circ < j < 90^\circ$, where $i$ is the inclination of the rotation axis towards the observer, and $j$ is the rotational angle  perpendicular to $i$ and on the horizontal plane which contains the direction of line of sight. When applying the stellar inclination, we rotate the star under the rotational coordinate transformation by $i$ first, then by $j$.}

Starspots are then introduced to the stellar surface by displacing warm photosphere with cooler (albeit nonzero) emitting regions, making the star appear redder and fainter.  The radius is kept fixed to the \citet{Baraffe2015} models, so the total luminosity of the star is less than a spot-free stellar surface. Diminishing the emergent luminosity creates an artificial mismatch between the internal energy production from contraction, fusion and the total luminosity emitted through the stellar surface.  This mismatch would ordinarily lead to radius inflation, which has recently been observed in the Pleiades cluster \citep{Somers2017, Jackson2018} and in rapidly rotating M dwarfs \citep{Kesseli2018}, with stars outsizing predictions by about $10-18\%$.  We assume for now that radius inflation would translate to isochrones but would not introduce additional scatter on H-R diagram.  Faculae, accurate limb-darkening effect, and warm chromospheric activity are not considered in these simulations.
 
The emission from both cool spots and the photosphere are calculated from the BT-Settl models  (version CIFIST2011\_2015, \citet{Allard2014} with solar abundances from \citet{Caffau2011}). The synthetic spectra are generated based on the $T_{\rm phot}$, $T_{\rm spot}$ and gravity from the \citet{Baraffe2015} pre-main-sequence evolution grid.  The stellar emission is hence defined by combining two temperature components modulated through the spot filling factor. The flux in the Johnson's $B$, $V$, $I$, and 2MASS-$K_s$ photometric bands in any single epoch is calculated by integrating the synthetic spectra over generic filter transmission curves \citep[from SVO service;][]{Rodrigo2012} in flux space ($\rm W/(m^2\mu m)$).

\begin{figure}[!t]
\centering
\includegraphics[width=3.4in,angle=0]{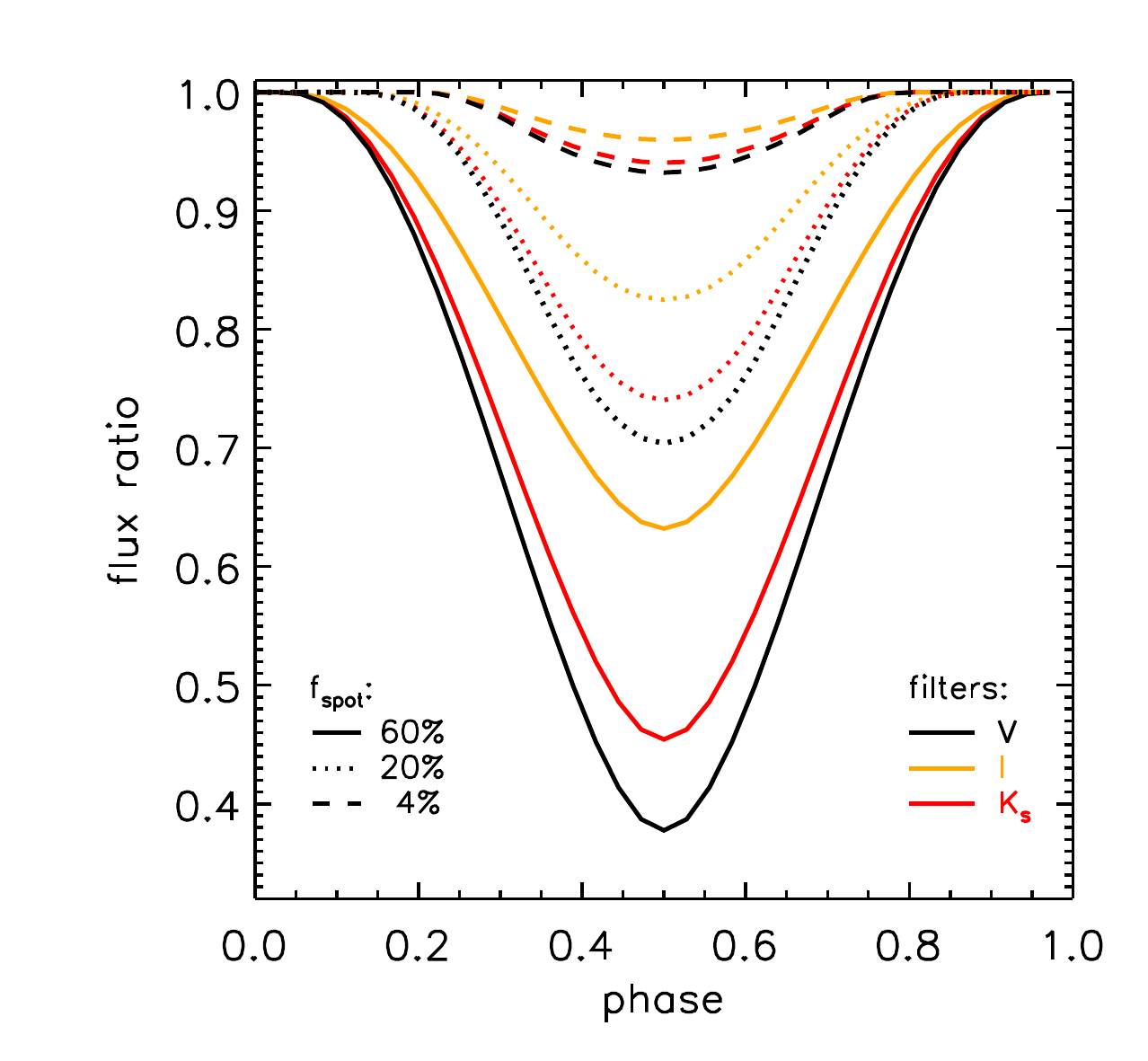}
\caption{Lightcurves modulated by a single spot with spot filling-factors ($f_{\rm spot} = $4, 20 and 60\%) and a fixed spot temperature $T_{\rm spot} = 3500$ K on a  K-type star ($T_{\rm phot} = 4000$ K). Three optical bands are adopted in the lightcurve, shown by different colors ($V$: black, $I$: orange, $K_s$: red). As the definition of spot coverage is $S_{\rm spot} / S_{\star}$, the 60\% coverage of a single spot means more than half of the star is covered by spot where $F_{min} \sim (T_{\rm spot}/T_{\rm phot})^4$.}
\label{fig:single}
\end{figure}

\begin{figure}[!t]
\centering
\includegraphics[width=3.4in,angle=0]{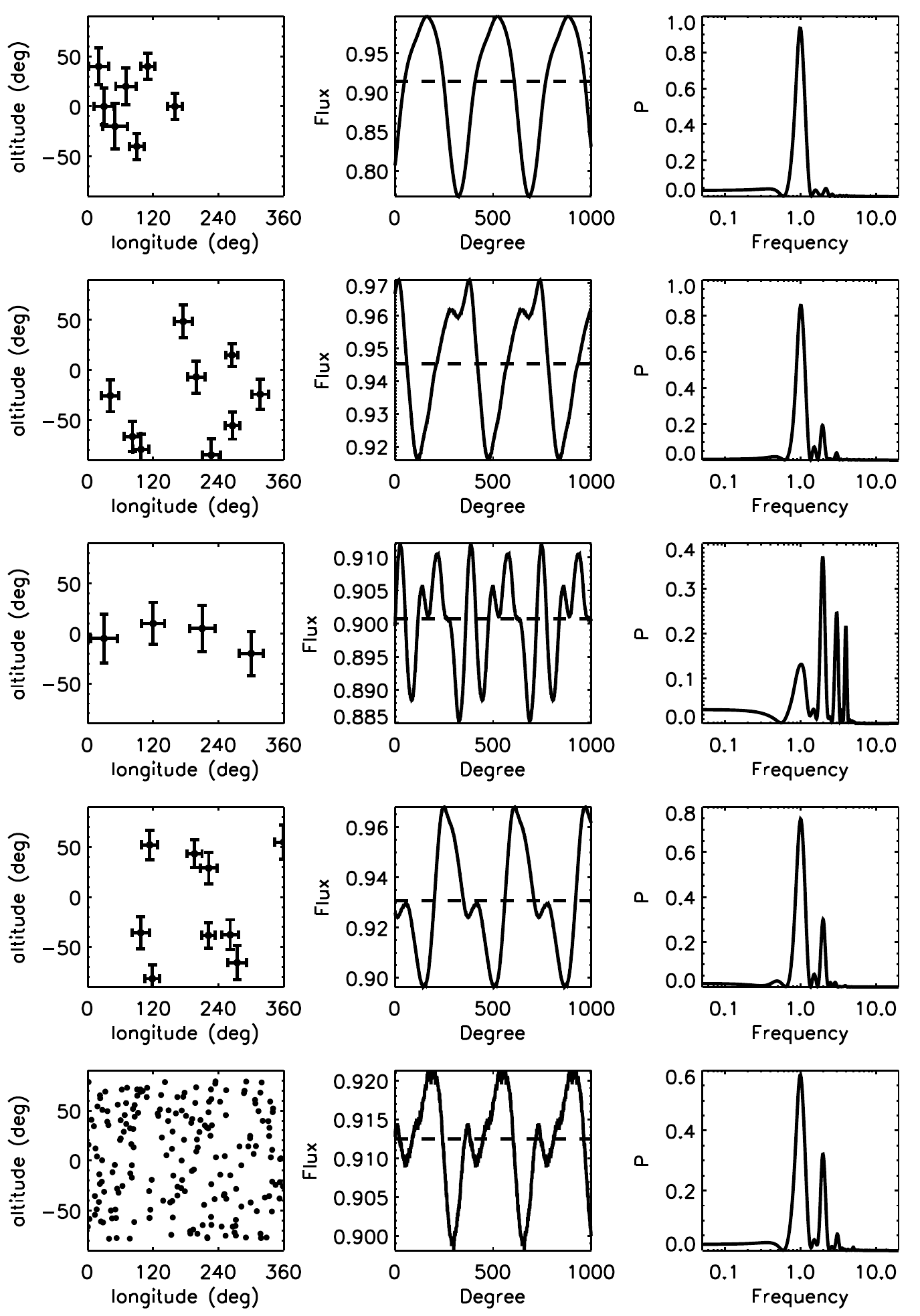}
\caption{From top to bottom, four examples of simulated ``multiple spot'' distribution and one example of `small spots' with the same parameters ($T_{\rm eff} = 4000$ K, $T_{\rm spot} = 3500$ K, $f_{\rm spot} = 10\%$).  Each horizontal panel shows spot distribution, lightcurve and periodogram. {\it Left:} Spot distribution on the ($\theta, \phi$) map. The error bars represent the semi major axises of the spots. {\it Middle:} Simulated lightcurve in unit of flux ratio against the photosphere through the {\it V}-band. The photospheric flux is set to 1.0. The horizontal dashed lines represent the average value in each figure. {\it Right:} Periodogram of the lightcurves calculated from the generalized Lomb-Scargle program. The stellar rotation frequency is normalized to 1.0.}
\label{fig:example}
\end{figure}

\subsection{Spot Configurations}

In this section, we introduce how cool spots are simulated on stars. First, the stellar surfaces are constructed by two-dimensional maps with latitude $\theta \in (-90, 90) $ and longitude $\phi \in (0, 180)$ with a resolution of 1 deg on each angular axis. Then, spots are introduced with ranging size and temperature. The stellar rotation is simulated by moving pixels step by step along the longitude axis on the ($\theta$, $\phi$) map. The ($\theta$, $\phi$) map is then converted to an observed surface map $(\theta_{obs}, \phi_{obs})$ via rotational coordinate transformation according to inclination angles. For an inclination of $90^\circ$, each point on the stellar surface is visible for half of the revolution, while for an inclination of $0^\circ$, only one hemisphere is ever visible and the lightcurve would be constant.

The simulated spots on the ($\theta$, $\phi$) plane are controlled by five variables: the number of spots on stellar surface ($N_{\rm spot}$), and the size ($S_{i}$), the temperature  ($T_{\rm spot, \,i}$), as well as location ($L_{\rm spot} (\theta_i, \, \phi_i) $) of each individual spot . The total spot filling factor $f_{\rm spot}$ on Cartesian coordinates constrains the $N_{\rm spot}$ and $S_{i}$, reducing the number of independent variables by one,
\begin{equation}
f_{\rm spot} = \big( \sum_{i = 0}^{N_{\rm spot}} S_{i} \big) /S_{\star},
\end{equation}
\noindent where $S_{\star}$ represents the surface area of the star. The range of spot coverages in the simulations is $f_{\rm spot} \in (30 - 50) \%$  for late type stars ($T_{\rm phot} < 3800 K$) and (1--40)\% for warmer stars, notionally consistent with the measurements from the TiO band absorption seen with  LAMOST \citep{Fang2016}. In our initial simulations,  $T_{\rm spot}$ is a free parameter between 2500 K and 50 K below the $T_{\rm phot}$ for each test star, according to previous spot temperature measurements summarized in \citet{Berdyugina2005,Strassmeier2009}.

We assume three different spot morphologies: (a) a single giant spot, (b)  multiple spots, with 5--10 medium size spots randomly distributed on the stellar surface, and (c) small spots, with 100--200 spots that cover the stellar surface. {  The last case is comparable to the solar spots, which are dominated by small-size spots ($\sim 10^{-6} S_{1/2 \odot}$) and follow a log-normal distributions \citep{Bogdan1988, Solanki2006review} as}
\begin{equation}
dN / dS \propto \exp [- (\ln S - \ln \bar{S})^2/ \ln \sigma^2_{S} ],
\end{equation}
\noindent where $\bar{S} = f_{\rm spot} / N_{\rm spot}$ is the mean value of the spot sizes, and $\sigma_{S}$ is the standard deviation of spot sizes, with $\sigma_{S} = 10 $ ppm from estimates of magnetic active stars  \citep{Solanki1999, Solanki2004spot, Barnes2011}.

{  In our simulations, the shapes of each individual spot are circular for the ``single giant spot'' and ``small spots'' configurations before placing them onto the stellar surface. For the ``multiple spots'' configuration, each spot is an ellipse with eccentricity $e < 0.5$. The axes of the ellipse are along the longitude and latitude directions.  The upper limit of eccentricity is set as 0.5 to prevent highly elongated spots. The extremely elongated spot along a large range of longitude is rarely detected on the sun and is not considered in our simulations. The key parameters that modulate the lightcurves are the spot coverage and their general distribution on the stellar surface. The shapes of individual spots do not affect the large structures of the lightcurves. Since, some spots overlap on the stellar surface, the total spot coverage is slightly smaller than the given $f_{\rm spot}$ when counting these overlapped regions.} 

Figure \ref{fig:example_morphology} shows four maps, one of a pure photosphere and three examples of spot morphologies.  Morphologies that are azimuthally symmetric, including polar spots and rings, are not shown and would not contribute to any  variability.  Such longitudinally-symmetric structures would produce color and brightness offsets from a spot-free photosphere detectable in spectral decomposition approaches and are discussed in \S 3.4.

Finally, the observed stellar emission ($F_{\rm obs}$) is calculated following the ``double-cosine'' rule \citep{Rackham2018} for each pixel as,
 \begin{equation}
F_{\rm obs} = \Big( \sum_{\phi = - 90}^{+90} \sum_{\theta = 0}^{180} f({\theta,\phi}) \cos(\theta - 90^\circ)\cos(\phi) \Big) \times F_{\rm phot},
 \label{eq:1}
 \end{equation}
\noindent where $ f({\theta,\phi}) \in (0, 1)$ is the ratio of spot-blended stellar emission ($F_{\rm spot}$) and the photospheric emission ($F_{\rm phot}$) of each pixel. Lightcurves are generated by recording $F_{\rm obs}$ step by step with stellar rotations. {  The darkening process in this paper is simulated by this ``double-cosine'' rule. The rule itself is a geometric effect when converting the flux from the $(\theta,\phi)$ map. This simplified limb-darkening does not consider the radiative transfer within the stellar atmosphere and hence does not change between the photometric temperatures. Since the core argument of this paper focuses on overall color and magnitude changes introduced by cool spots, rather than on the fine structure of the lightcurves, we choose to keep this ``double-cosine'' rule as an approximation of the limb-darkening effect.}

In the following analysis, two values are defined to describe the variations of lightcurves in linear space. $\Delta F_{\rm LC}$ as the 10 to 90 percentile amplitude of the simulated lightcurve, while $\Delta F_{\rm spot}$ is the amplitude of minimum observed stellar flux compared to the flux emerging from an entirely spot-free stellar photosphere with otherwise identical properties. The term $\Delta F_{\rm LC}$ is defined similarly as the photometric amplitude in Figure \ref{fig:k2_lamost}, while $\Delta F_{\rm spot}$ resembles $\Delta F_{\rm spec}$ defined in Equation 1.

\begin{figure}[!t]
\centering
\includegraphics[width=3.5in,angle=0]{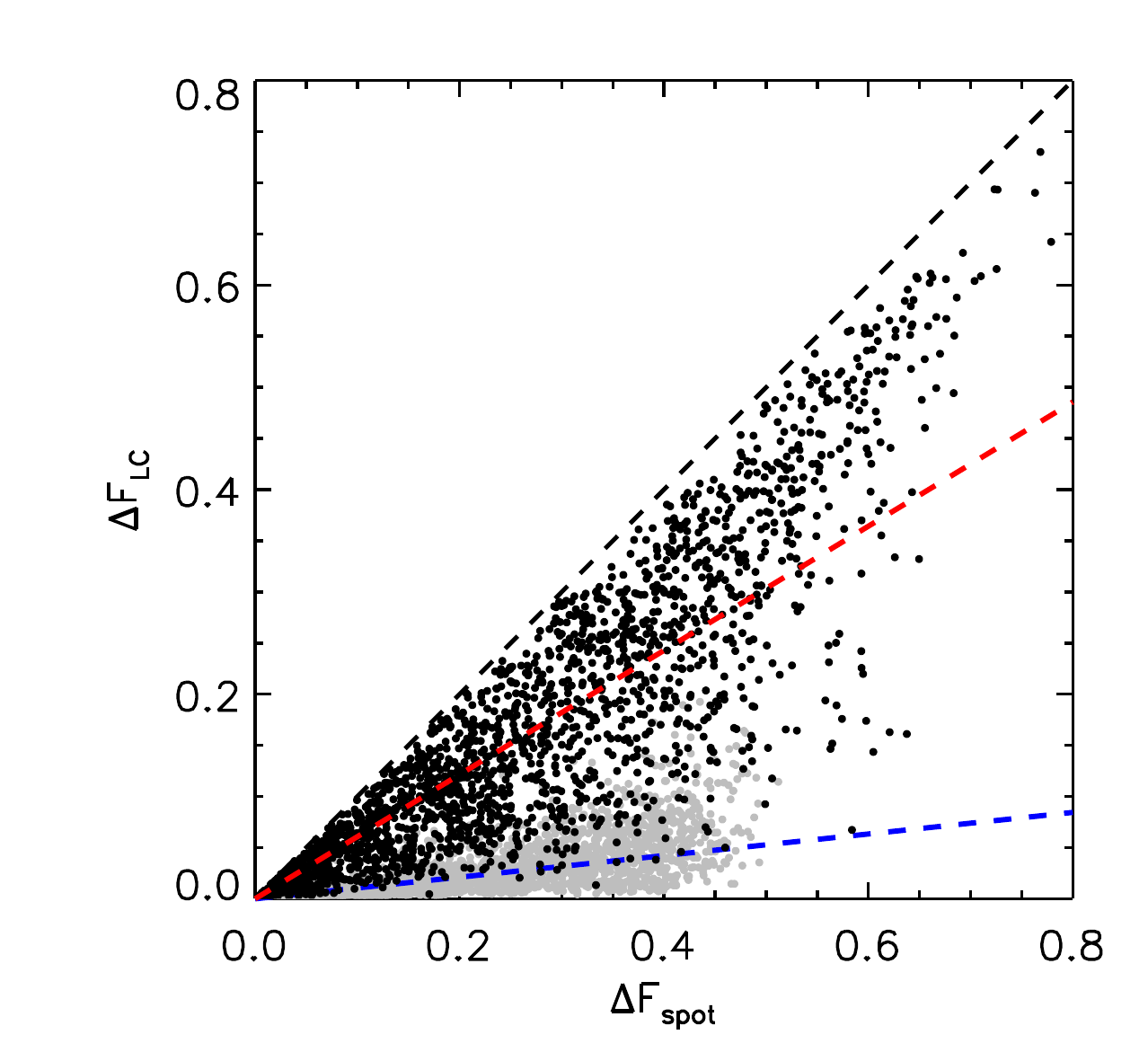}

\caption{A comparison between the spot covering amplitude, $\Delta F_{\rm spot}$, and the observed variation amplitude in $V$-band. The ``multiple spot'' case is shown by the black dots, while the ``small spots'' are shown by grey dots. The black and red dashed-lines represent the detection rate $\zeta = \Delta F_{\rm LC}/\Delta F_{\rm spot} =$ 1.0 and 0.53, and the blue dashed line is $\zeta = 0.08$. }
\label{fig:multi}
\end{figure}

\subsection{Lightcurve analysis}

The simulated lightcurves have amplitudes and power spectra that depend on how spots are distributed on the star. The lightcurves simulated with the simplest spot morphology, the ``single spot'' case, are shown in three photometric bands in Figure~\ref{fig:single}.  At a fixed spot temperature, the spot filling factor ($f_{\rm spot}$) determines the photometric amplitudes. Meanwhile, with similar $f_{\rm spot}$ and $T_{\rm spot}/T_{\rm phot}$, spots create larger luminosity variation on cooler stars than warmer stars. Among the lightcurves, the $V$-band always show larger variation amplitude as an indicator of color dependency (see \S 4.2 for more information).

\begin{figure}[!t]
\centering
\includegraphics[width=3.5in]{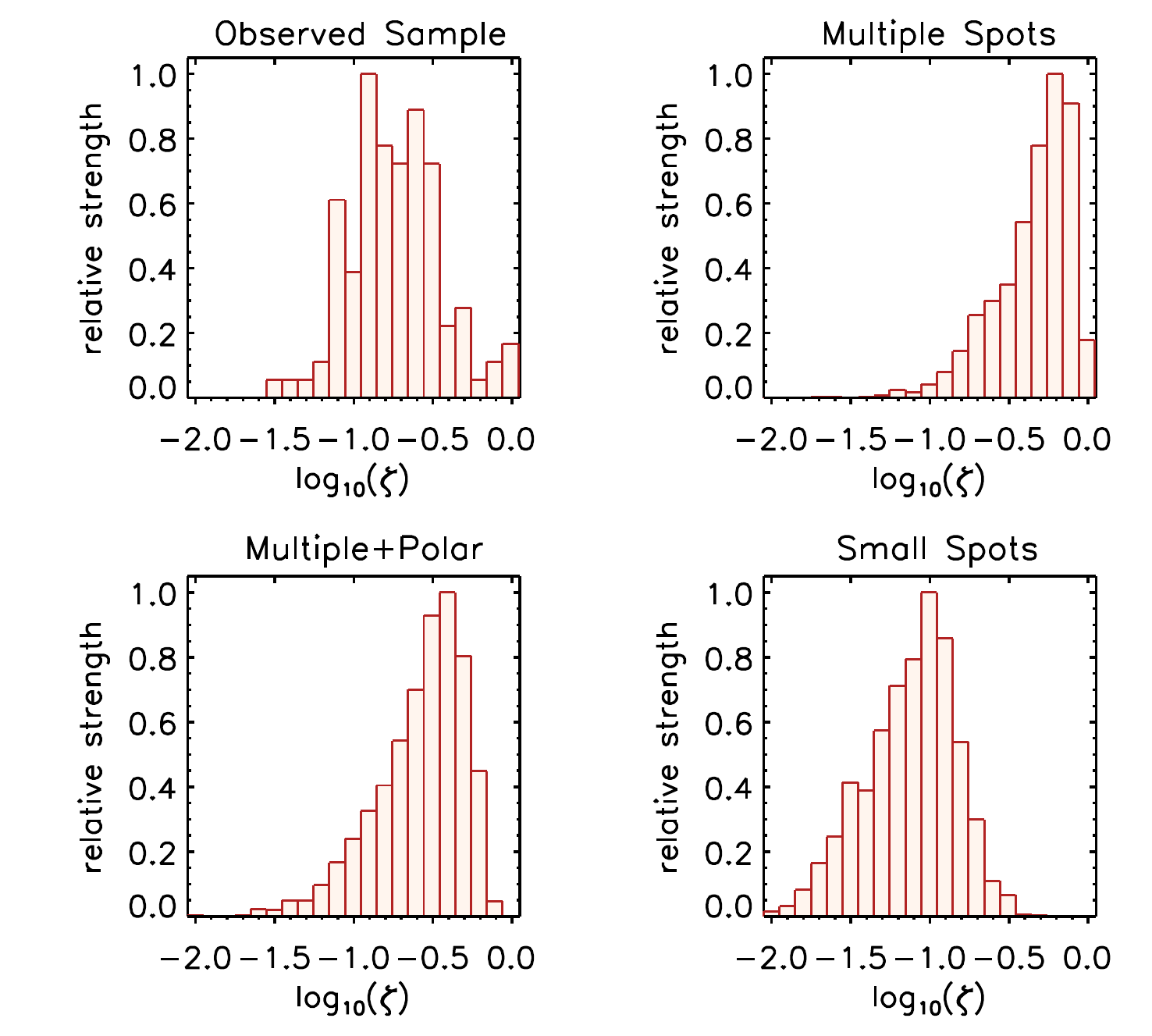}
\caption{ {  Histograms of $\zeta$-values ($\Delta F_{\rm LC}/\Delta F_{\rm spot}$)} from K2 and  LAMOST observations (Figure \ref{fig:k2_lamost}) and three simulations (multiple spots, multiple + circumpolar spot, and small spots).}
\label{fig:hist_k}
\end{figure}

\begin{figure*}[!t]
\centering
\includegraphics[width=7.in, angle=0]{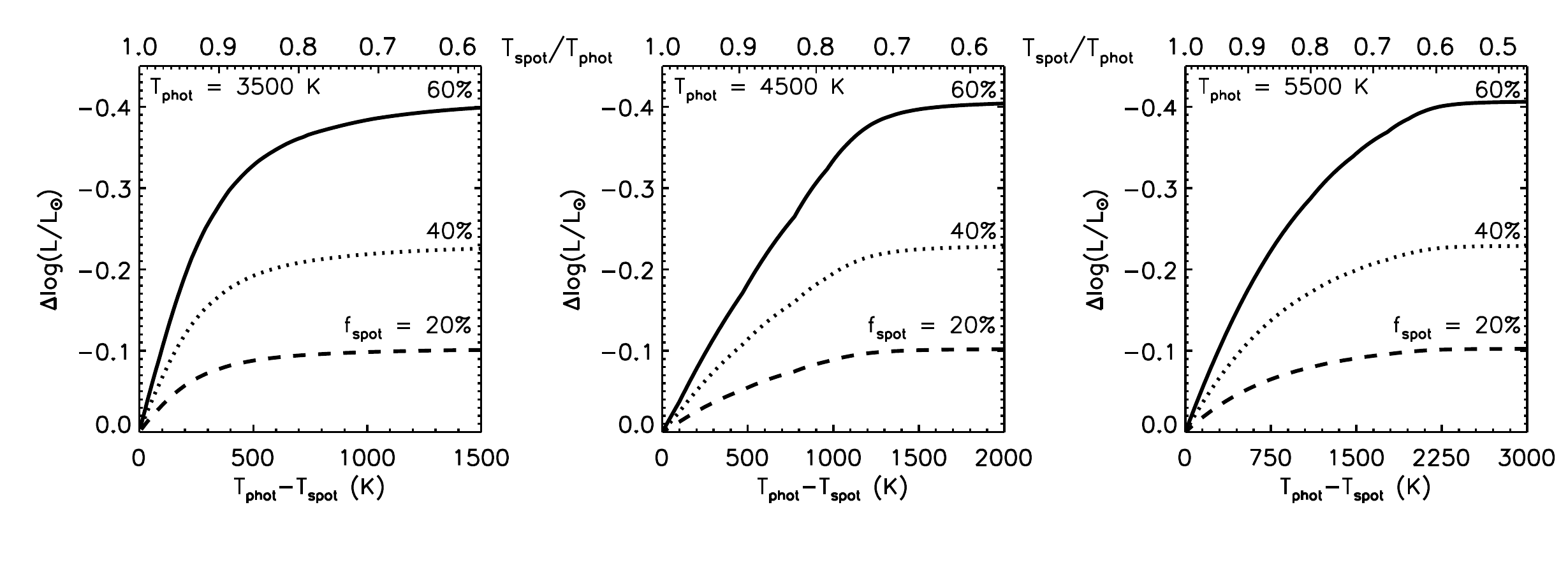}
\caption{The luminosity variation of spot-covered stars versus spot temperature. From left to right, each panel represents photospheric temperature as 3500, 4500 and 5500 K. Three spot coverages are shown within each panel, as $f_{\rm spot} = 20\%, 40\%$ and 60\%, from bottom to top, marked by dashed, dotted and solid lines, respectively.}
\label{fig:lum}
\end{figure*}

{  Although the ``multiple spots'' morphology is geometrically more complicated than the ``single spot'' case, the methodology of generating lightcurves is the same. As described in Section 3.2, multiple spots are put onto each simulated stellar surface with a range of spot parameters, and the lightcurves are generated through synthetic spectra and stellar rotation.}  Figure \ref{fig:example} presents a few examples of spot configurations with their output lightcurves. In general, the distribution of spots controls the shapes and the varying amplitudes of the observed lightcurves. More azimuthal clustering of the spots generates lightcurves with larger amplitudes, as the top panel of Figure \ref{fig:example}.
When spots are distributed symmetrically in the longitudinal direction, as in the third panel, the photometric variability is low because one spot becomes visible as another spot rotates to the unseen side of the star.  The bottom panel of Figure~\ref{fig:example} presents an example of the ``small spots'' case with 191 randomly distributed spots on the stellar surface.  The small asymmetry in the spot distribution leads to a periodic synthetic lightcurve with a small amplitude, here 2\% in $V$-band, which is the smallest among all examples.

The frequency analysis of synthetic lightcurves (the right column of Figure~\ref{fig:example}) is obtained using the generalized  Lomb-Scargle periodogram \citep{Zechmeister2009}, which calculates the power over a range of frequencies, assuming that the lightcurve is well explained by sinusoidal functions. In all cases, the input frequency is the smallest frequency rather than the highest peak.  For most simulated lightcurves, the frequency analysis yields peaks in the power at several different periods because of coincidental resonances.  The identification of multiple periods in lightcurves reproduces the general behavior of the K2 lightcurves of the Pleiades analyzed by \citet{Rebull2016b}. 
For K2 lightcurves with multiple peaks in the power spectra, the lightcurve amplitudes are generally smaller than stars with single periods, which is well explained with our simulations.  Multiple periods may also be detected in unresolved multiple systems.

\subsection{Comparing detection ratio between morphologies}

To quantify the relationship between the spot distribution and the amplitude of lightcurve, we define a parameter $\zeta = $ $\Delta F_{\rm LC}$ / $\Delta F_{\rm spot}$, as the ratio between the 10--90 percentile amplitude of the lightcurve divided by the amplitude of the minimum stellar flux against a pure photosphere.  The value of $\zeta$ quantifies the symmetry in the distribution of spots on the stellar surface by the shape and depth of lightcurves, ranging from 0 for symmetric distributions, such as a ring or polar spot, to 1 for highly asymmetric distributions, such as a single spot.  This ratio also depends on inclination, since viewing angles that are closer to pole-on reduce variability and lead to a lower value of $\zeta$.

Figure~\ref{fig:multi} compares the amplitude of the lightcurve to the spot coverage for our simulations.  For the ``multiple spots'' case, the median amplitude ratio $\zeta$ is 0.53. Simulations with $\Delta F_{\rm spot} > 50\%$ require clustered spots, leading to large $\zeta$ in all such cases.  The median $\zeta$-value represents that the variation detected by the lightcurve is only two-thirds of the realistic brightness changing on the star due to the symmetric distribution of spots. The grey dots in Figure~\ref{fig:multi} present the ``small spots'' case with smaller amplitude as well as $\zeta$-values. In other words, if explained by small spots, a $1\%$ amplitude in a K2 lightcurve corresponds on average to a $10\%$ coverage fraction of starspots.  

We add polar spots to these simulations, referred to here as ``multiple+polar'' simulations, to approximate the large high-latitude spots that are often detected on magnetically-active stars \citep{Donati2008, Barnes2015}. The sizes of the polar spots shown in the ``multiple+polar'' configuration are fixed to 50\% of the total spot coverage on each star. Figure~\ref{fig:hist_k} summarizes the $\zeta$-values in log space for three simulated morphologies (multiple, multiple+polar and small spots) and for the observational results of K2 and LAMOST presented in Figure~\ref{fig:k2_lamost}. The distribution of observed $\log_{10}(\zeta_{\rm obs})$ is located between the ``small spots'' and ``multiple+polar'' case, while the ``multiple spot'' case results in the largest detection ratio significantly offset from the observed sample. {  By comparing the ``multiple+polar'' and ``multiple spot'' cases, the median detection ratio ($\zeta$) is decreased by 0.22 dex when 50\% of the spots are located at the polar region. By enlarging the sizes of polar spots, the median detection ratio becomes smaller, and reaching the observed value (K2/LAMOST) when 70\% of the spots are polar spots.} Meanwhile, the detection ratio of the ``small spots'' morphology is an average of 0.1 dex lower than the observed sample. The observed distribution of detection ratios can be explained by a highly-symmetric spot morphology, like the ``small spots'' configuration. The detection ratios for different morphologies are listed in Table \ref{tab:k-value}.  

The simulation results confirm our intuition that the bulk of starspots do not induce temporal modulation because of longitudinal resonances.  One spot exits the observable stellar hemisphere as another spot enters, balancing the loss of diminished flux. Realistic spot distributions are likely more complicated than our assumptions and likely include large symmetrically distributed spots, or not-randomly distributed small spots associating with magnetic field activities. The spot symmetry suppresses the temporal modulation amplitude in the Kepler/K2 lightcurves.

\begin{table}[!b]
\centering
\caption{$\zeta$-value measured from observation and simulations}
\begin{tabular}{lcccc}
\hline
\hline
Group & Band & median $(\zeta)$ & standard deviation ($\sigma({(\zeta)})$) \\
  \hline
  Observation Sample & {\it Kepler} & 0.18 & 0.23\\
 Multiple   & $V$ & 0.53 & 0.23\\
  Multiple+polar & $V$  & 0.31 & 0.16\\
  Small spots & $V$ & 0.08 & 0.06\\
  \hline
  \hline
  \label{tab:k-value}
\end{tabular}
\end{table}

\section{The observational effects of starspots}
Spots affect our ability to accurately measure the radius and effective temperature of the star.  In this section, we simulate color-magnitude and H-R diagrams for different spot parameters to evaluate how spots will change the measured properties of stars in a cluster.  A specific ``multiple spots'' configuration is adopted in this section since the following discussions are independent to the spot distribution and stellar rotation. We first evaluate the changes in luminosities and colors caused by spots and then create populations to determine the effects on color-magnitude and H-R diagrams.

\begin{figure*}[!t]
\centering
\includegraphics[width=6.5in, angle=0]{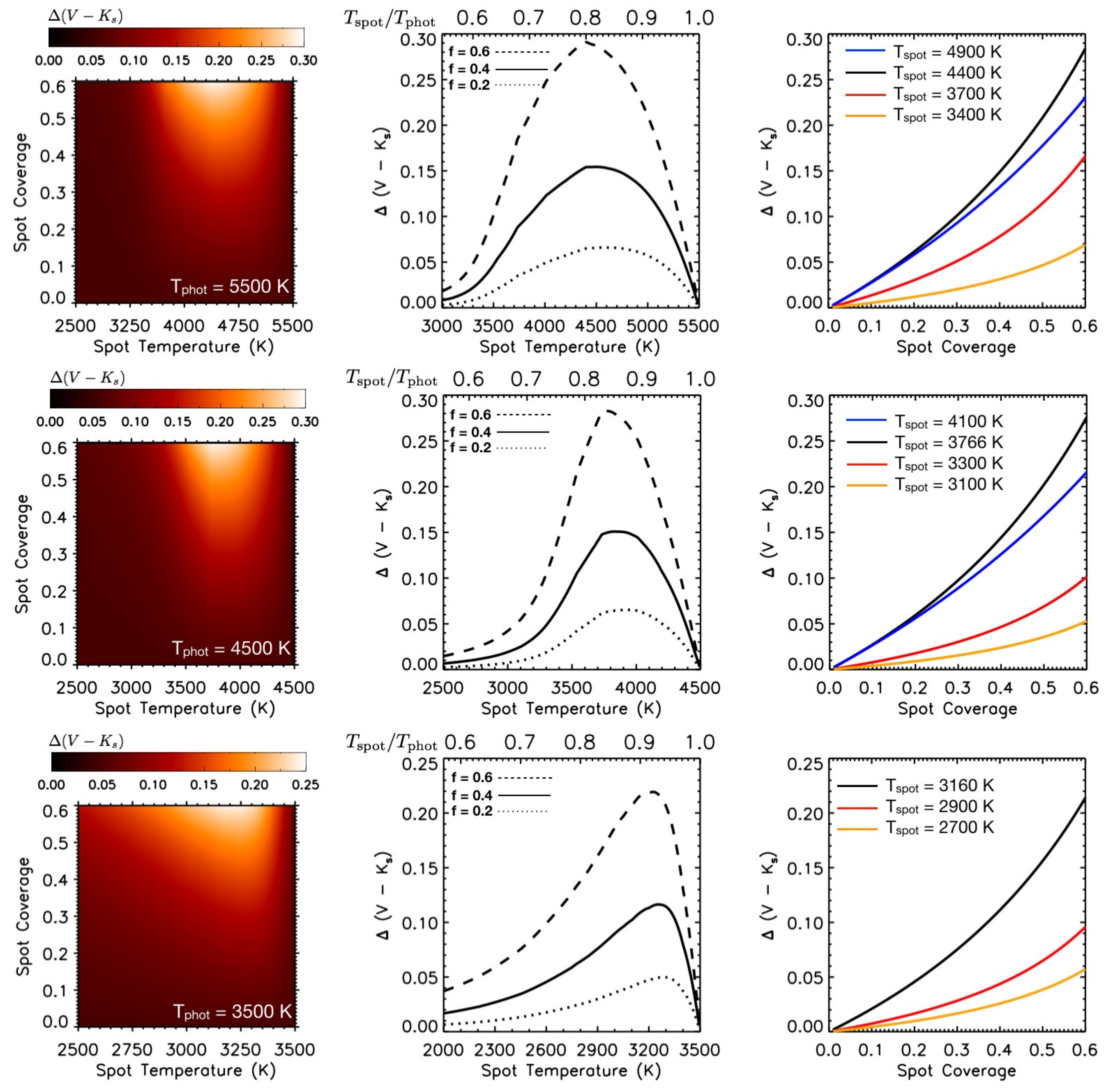}
\caption{The color variation, $\Delta (V - K_s)$, introduced by spot modulation on stellar surface with different spot filling factor ($f_{\rm spot}$) and spot temperature $T_{\rm spot}$ {  under the `multiple spot' configuration}, for photopsheric temperatures of 5500, 4500, and 3500 K.  Left panels are 2-D figures of $\Delta (V - K_s)$ in the spot parameter space. The values of $\Delta (V - K_s)$ are shown by the color (see the color bars). The middle and right panels are 1-D plot of $\Delta (V - K_s)$ versus different spot parameters. Detail parameters are labeled in the figure.}
\label{fig:v-k}
\vspace{0.3cm}
\end{figure*}

\subsection{Luminosity variations introduced by spots}

The first consequence of adding cool spots on stellar surface is the decay of observed stellar luminosity. At a fixed radius, spots decrease the stellar luminosity by
\begin{equation}
L_\star = L_{\rm phot} \times (1 - f_{\rm spot}) +  L_{\rm spot} \times  f_{\rm spot},
\label{eq:5}
\end{equation}
where $L_{\rm phot}$ and $L_{\rm spot}$ depend on the respective temperatures. 
Figure~\ref{fig:lum} presents the luminosity variation ($\Delta \log(L/{\rm L_\odot}) = \log(L_{\rm phot}/{\rm L_\odot}) - \log(L_\star/{\rm L_\odot})$) versus spot temperature ($T_{\rm spot}$) on three example stars, with  photospheric temperature ($T_{\rm phot}$) of 3500, 4500 and 5000 K. The luminosity variation is sensitive to spot coverage and asymptotically approaches the maximum when $T_{\rm spot}/T_{\rm phot}$ is lower than 0.6. The maximum luminosity variation occurs when no flux is emitted from the spot region, or so called ``non-emitting spot'' case, resulting in  $\Delta \log(L/{\rm L_\odot}) = \log(1-f_{\rm spot})$.
 
In our simulations, the stellar luminosities are calculated by integrating the flux from BT-Settl models over the ($\theta,\phi$) space, as shown in Equation \ref{eq:1}. When considering the fainter photospheric region at the edges (see Figure \ref{fig:example}) and limb darkening effects, the realistic luminosity variation would be slightly larger than the calculation above. However, the basic relationships between $\Delta \log(L/{\rm L_\odot})$ and spot parameters are the same.

\begin{figure*}[!t]
\centering
\includegraphics[width=6.3in, angle=0]{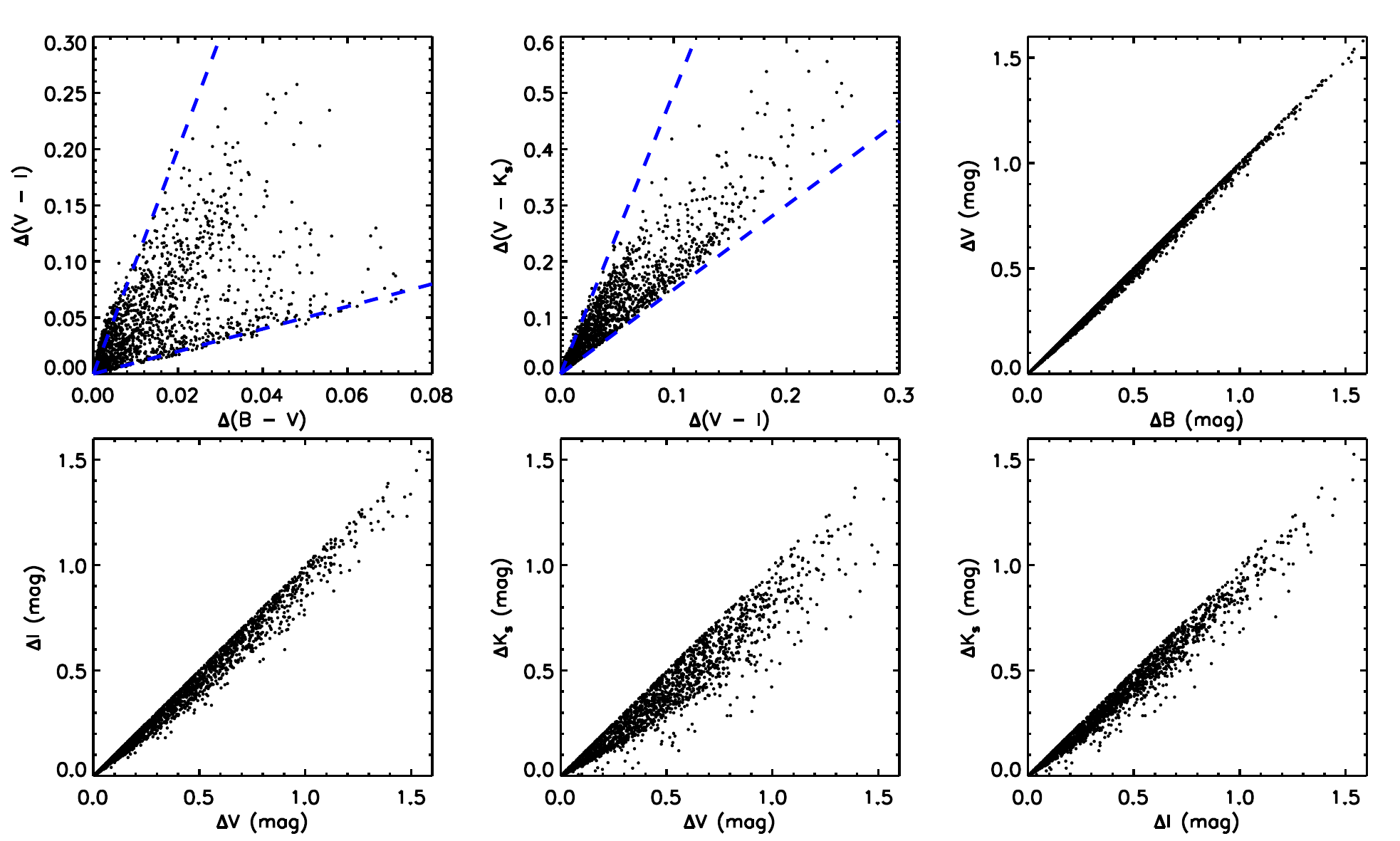}
\caption{Comparisons among the changes of stellar brightness caused by spot modulation in $B$, $V$, $I$ and $K_s$-bands. The blue dashed lines in the upper left panel represent $\Delta(V - I) = \Delta (B - V)$, and $\Delta (V - I) = 10 \Delta(B - V)$. Similarly, the blue dashed lines in the upper middle panel shows $\Delta(V - K_s) = 1.5 \Delta (V - I)$ and  $\Delta(V - K_s) = 5 \Delta (V - I)$}
\label{fig:delta_amp}
\end{figure*}

 \subsection{Color variation introduced by spot parameters}
In this section, we reveal the observational outcomes of various spot parameters, $f_{\rm spot}$ and $T_{\rm spot}$, on obtained stellar color. To minimize the contribution of the geometric distribution of spots, we only consider the ``single spot'' configuration here. As an example, the color variation between $V$ and $K_s$ bands are displayed in this section, as $\Delta (V - K_s)$, which represents the largest color change introduced by certain stellar and spot parameters throughout the stellar rotation. The $\Delta (V - K_s)$ is calculated from the synthetic lightcurves of two colors, and the results are displayed in Figure \ref{fig:v-k}. The synthetic color-color diagram is then compared to the observed colors of stars in the Pleiades for understanding the spot behaviors on cluster members.
 
The spot coverage dominates the variations of observed colors within certain ranges of $T_{\rm spot}$. The curves of $\Delta (V - K_s)$ versus $f_{\rm spot}$ (the right panels, Figure~\ref{fig:v-k}) are
 similar for different sets of photospheric temperatures. For instance, the maximum $\Delta (V - K_s)$ reaches 0.06, 0.15 and 0.28 mag for spot filling factor $f_{\rm spot} = $ 0.2, 0.4 and 0.6 when $T_{\rm phot} =$ 5500 K.

On the other hand, with a fixed spot filling factor,  $\Delta (V - K_s)$ always reaches a maximum at $T_{\rm spot}$ that is around 80--90\% of $T_{\rm phot}$.  At higher spot temperatures, the spot and photosphere colors are similar.  At lower spot temperatures, the spot flux becomes increasingly negligible and would cause the star to have a fainter luminosity without affecting the colors. The heuristic for $T_{\rm spot}$ scaling matches previous simulations of starspots, in which the choice of $T_{\rm spot}$ is either treated as a free parameter in a certain range \citep{Desort2007}, or as a fixed temperature difference against the photosphere in linear \citep{Barnes2011} or logarithmic space \citep{Rackham2018}.  
For realistic measurements of spot temperatures \citep[see Figure 7 and Table 5 from][]{Berdyugina2005}, cool stars ($T_{\rm phot} < 4500$ K) always have spot temperature as $T_{\rm spot} \sim 0.85 T_{\rm phot}$ that would result in maximum color variations. On the contrary, for warmer stars, only a few detections have such high $T_{\rm spot}$, while others are around $T_{\rm spot} \sim 0.7 T_{\rm phot}$.

\begin{figure*}[!t]
\centering
\includegraphics[width=2.31in, angle=0]{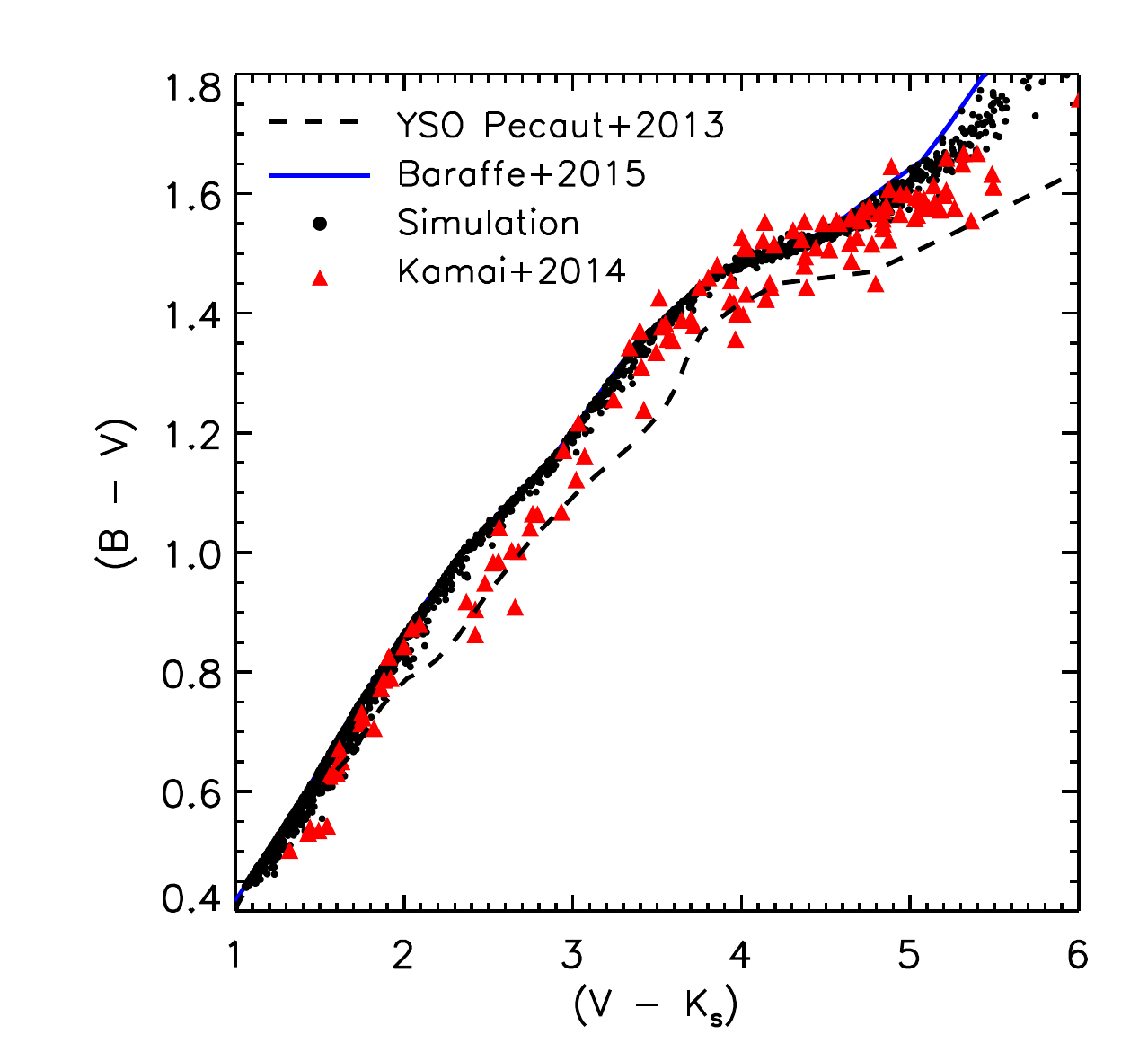}
\includegraphics[width=2.31in, angle=0]{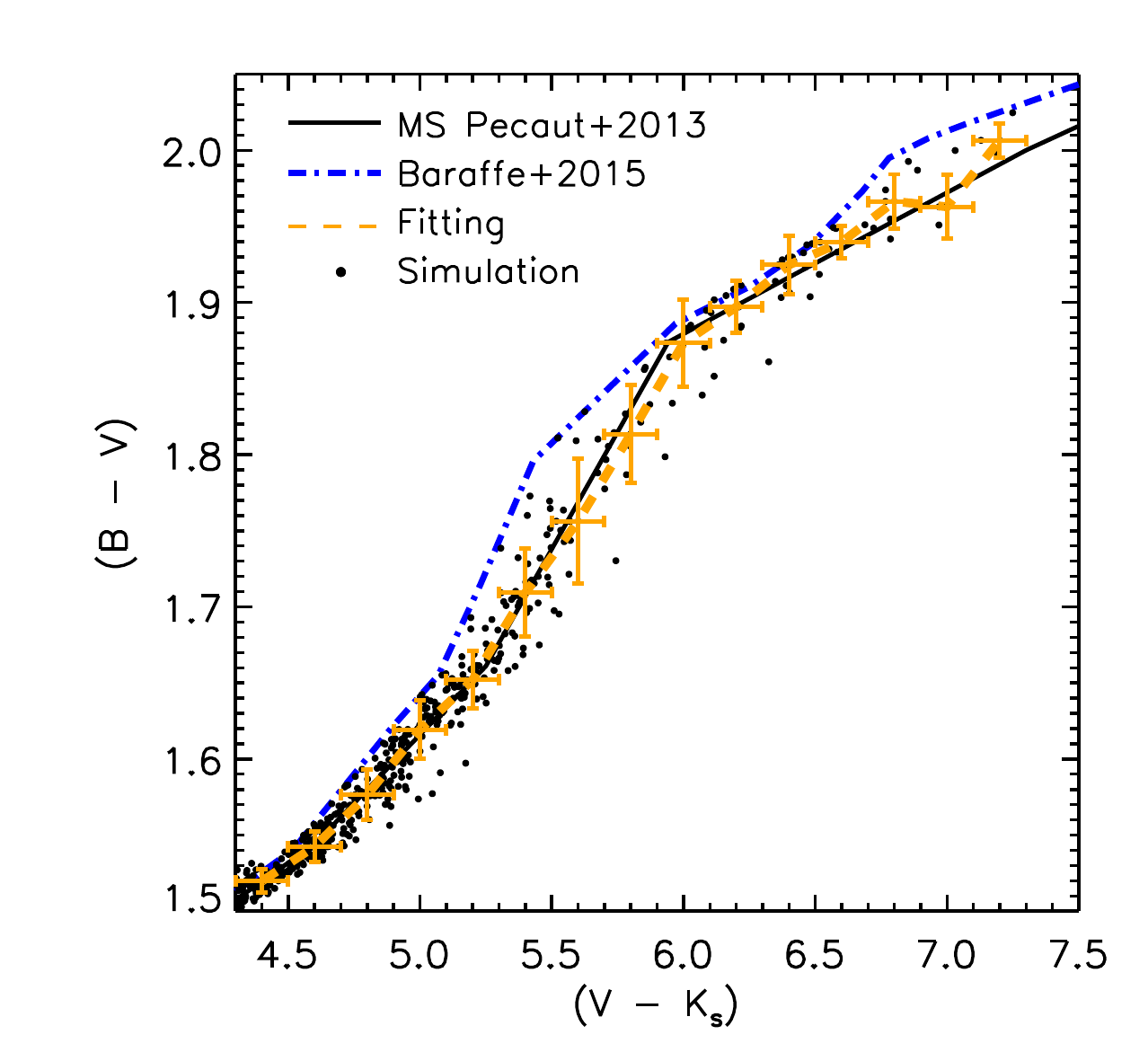}
\includegraphics[width=2.31in, angle=0]{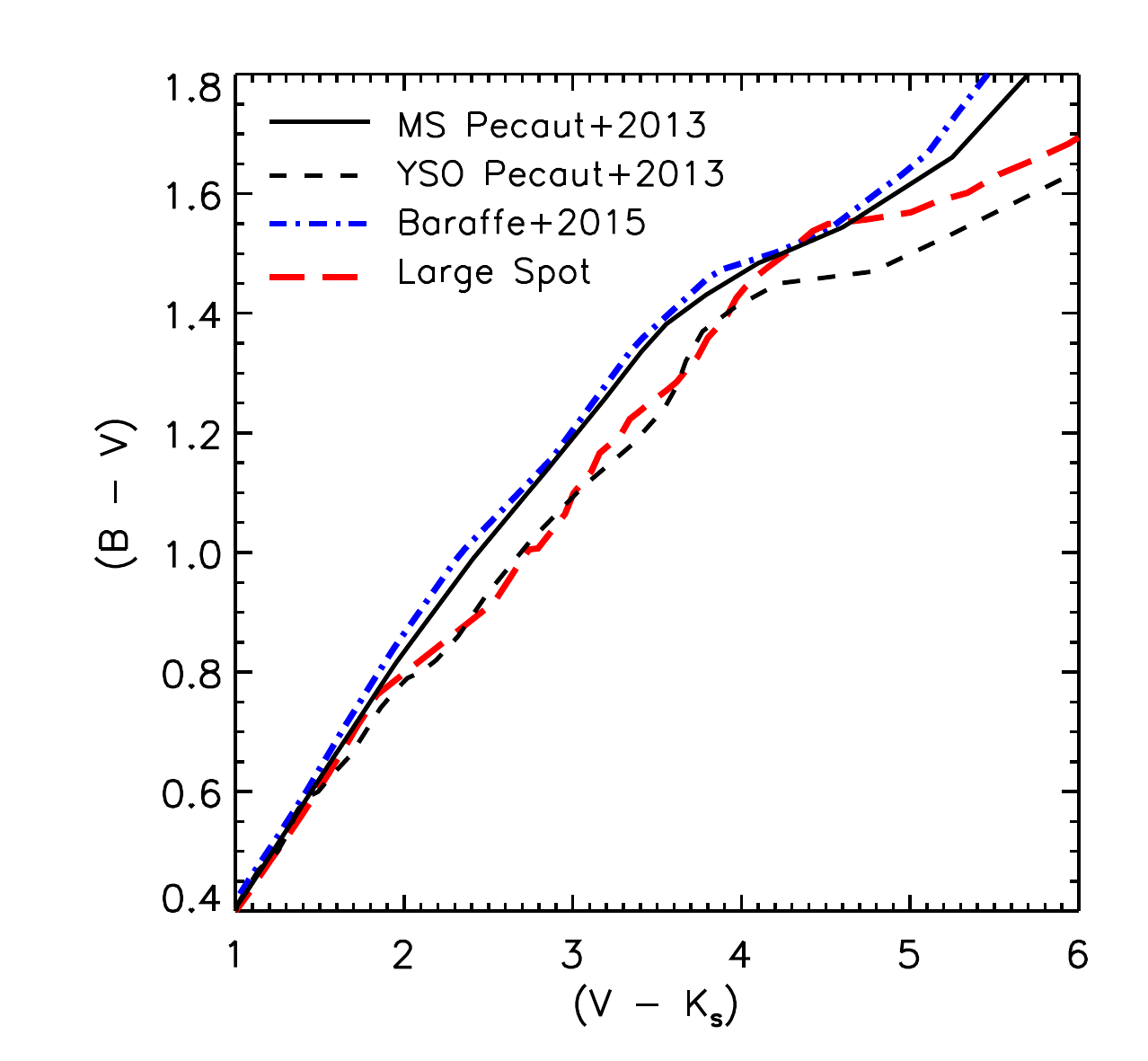}
\caption{Color-color diagrams in $(B - V)$ versus $(V - K_s)$ of observed (red triangles, \citealt{Skrutskie2006} and \citealt{Kamai2014}) and simulated (black dots) stellar samples with empirical and modeled isochrones. Three isochrones are shown in the figures, including empirical isochrones of main sequence and young stellar objects (solid and dashed black lines, \citet{Pecaut2013}), as well as the BHAC stellar evolution model (blue solid lines \citet{Baraffe2015}). The left panel shows a comparison between simulated and observed colors, while the middle panel provides a zoom-in value of low mass stars in the simulations. The orange dashed line in the middle panel is the binned simulated stellar color with error bars as the bin size (horizontal) and standard deviation within each bin (vertical). The right panel includes a calculated isochrone with 40\% spot coverage on hot stars ($(V - K_s) < 1.7$) and 90\% spot coverage on others, represented by the red solid line.}
\label{fig:color-color}
\end{figure*}

Our simulations of the stellar members in a coeval cluster compare the color and luminosity changes in different bands introduced by starspots (Figure~\ref{fig:delta_amp}). Since the peak of the black body of cold spot is located at the longer wavelength, the amplitude changes in bluer bands are larger than redder bands.  The relative color changes between $B$, $V$, $I$ and $K_s$-bands are smaller than the absolute magnitude decrease in each band. $\Delta (B - V)$ is the smallest among all the colors.

The effects of existing cool spots on changing the stellar loci on color-color diagrams are presented in Figure~\ref{fig:color-color}. The color-changing effects are shown by adding color variations on each sample in the simulation, shown in Figure~\ref{fig:delta_amp}, onto their photospheric colors calculated from the BHAC stellar models \cite{Baraffe2015}. 
Observed Pleiades members (see \S 5.2 for a description of Pleiades membership) are also shown for comparisons including $B$ and $V$-band photometry from \citet{Kamai2014} and $K_s$-band from 2MASS \citep{Skrutskie2006}. Large observed color spreads are seen on cooler stars indicating strong spots and plages events, which are generated from active stellar magnetic fields. 
 
The empirical colors from \citep{Pecaut2013} are also shown in Figure~\ref{fig:color-color} for comparisons.  A zoom-in view of low mass stars (e.g. $(V - K_s) > 4$) in the middle panel of Figure~\ref{fig:color-color} shows a notable discrepancy between the empirical main-sequence and photospheric colors.
Here, a spot-covered star cluster is generated by adding the simulated color changes from cool spots onto the modeled photospheric colors.  The binned simulated samples fit the empirical color of main-sequence stars, suggesting that the discrepancy between empirical colors of low mass main sequence stars and photospheric models might come from starspots.

The empirical color difference between young and MS stars found by \citep{Pecaut2013} might be related in part to the presence of cool spots. However, to recover this color difference, a 90\% coverage of cool spot is required for warm photosphere, or $(V - K_s) < 4$. For cooler stars, a bluer excess in $(B - V)$ is needed to fit the lower end of the isochrone. This might relate to faculae or chromospheric activity \citep{Kowalski2016} that most luminous in $B$-band. However, the 90\% spot coverage is abnormal even on the most magnetic active stars through an extreme spot coverage is indeed detected on a specific WTTS, LkCa 4 \citep{Gully2017}. In addition to stellar activities, gravity might be another cause of the color degeneracy but is beyond the scope of this paper.

\subsection{Footprints on color-magnitude and H-R diagrams}

Starspots cause observational consequences that bias the age determination of pre-main-sequence stars via stellar luminosity.  In this section, we discuss the footprints of starspots on the color-magnitude and H-R diagrams, which are sensitive to the spot parameters ($f_{\rm spot}$ and $T_{\rm spot}$). Since the geometric distributions of spots are not important when studying the observed ``snapshots'' on the color-magnitude and H-R diagrams, only the ``small spots'' configuration is displayed as other morphologies result in similar luminosity and color spread. All colors applied in the following analysis are based on the empirical colors of main-sequence stars \citep{Pecaut2013} plus simulated color changes. However, the luminosities are directly calculated from BT-Settl models following Equation~\ref{eq:5}.

Along with the simulations, six groups of spot parameters are defined to understand how spots with different parameters affect the stellar position for a star cluster on the H-R diagram (listed in Table~\ref{tab:par-group}). The parameters of Group 1 and 2 are selected to show the maximum luminosity and temperature spreads introduced by spots, where $f_{\rm spot} = 60\%$ is close to the maximum detection value in the Pleiades \citep{Fang2016}, while the $T_{\rm spot}$ is chosen as 0.6 $T_{\rm phot}$ and 0.85 $T_{\rm phot}$ according to our previous calculation in \S4.1 and \S4.2. The median and relatively low spot coverages ($f_{\rm spot} = 40\%$ and $20\%$) are applied in Group $3 - 6$.

\begin{figure*}[!t]
\centering
\includegraphics[width=4.5in, angle=0]{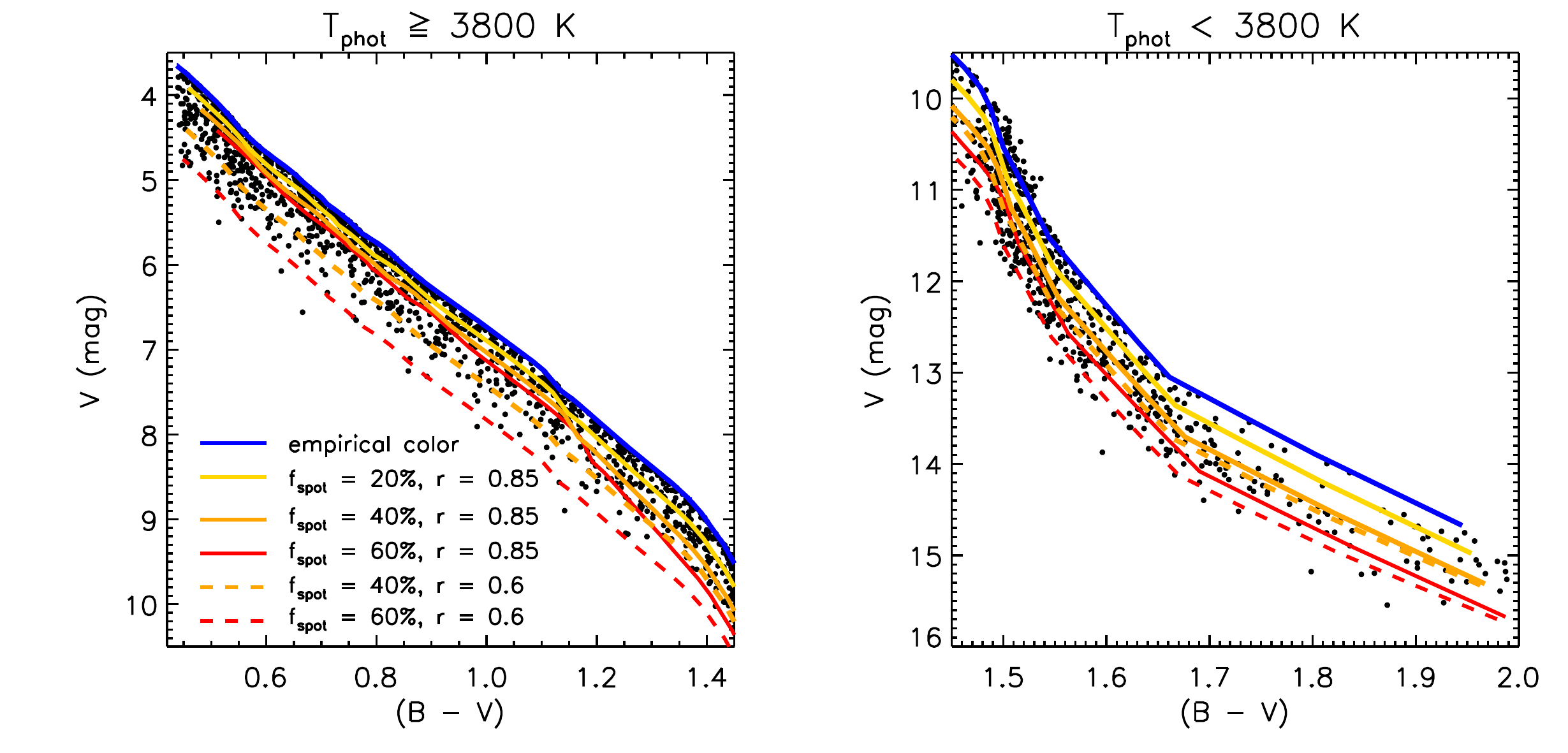}
\includegraphics[width=2.35in, angle=0]{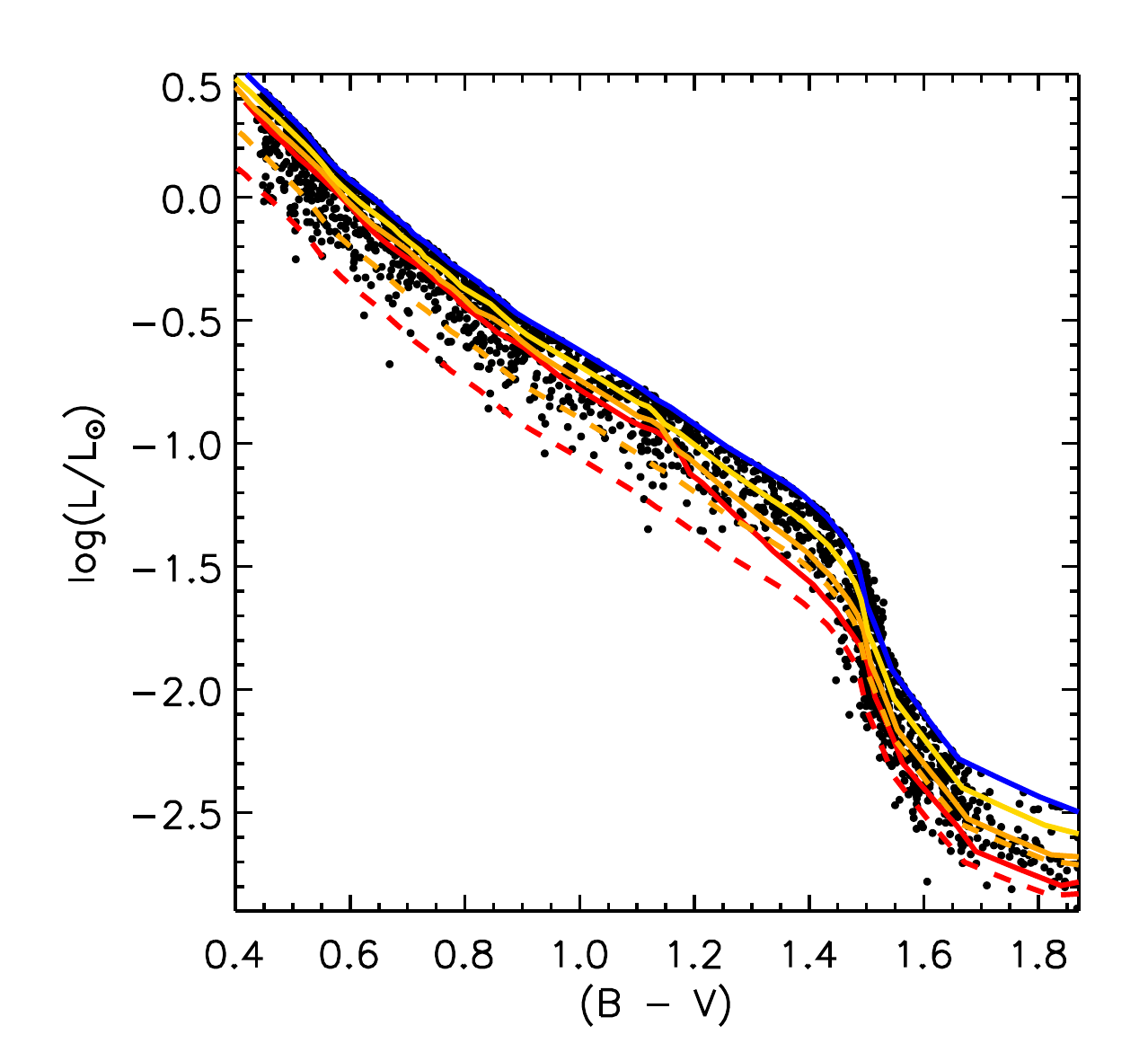}

\caption{Color-magnitude ($(B - V)$ versus $V$, left and middle panel) and a color-luminosity diagram (right panel) of simulated ``multiple spots'' covered star cluster. The blue lines in every panel are empirical colors \citep{Pecaut2013}.  Warmer stars ($T_{\rm eff} \geq 3800$ K) are shown in the left panel and cooler stars ($T_{\rm eff} < 3800$ K) are in the middle panel. The simulation results are shown by black dots, while the isochrones calculated from selected spot parameters (listed in Table \ref{tab:par-group}) are presented by lines with different colors and styles (see the keys in the left panel). Here $r = T_{\rm spot} / T_{\rm phot}$ is the temperature ratio between spot and photosphere. An H-R diagram is shown on the right panel, where the luminosity is calculated based on $V$-band magnitude and bolometric correction given by $(B - V)$ \citep{Pecaut2013}.}
\label{fig:cmd}
\end{figure*}

Color-magnitude ($(B - V)$ versus $V$) diagrams are shown in the left and middle panels of Figure \ref{fig:cmd} with simulated samples and five out of six models (Group 1--5, shown by colored solid and dashed lines\footnote{Since the results of Group 6 is very close to Group 5, we did not plot Group 6 in this section to keep Figure \ref{fig:cmd} simple}.). For warmer stars, the spot-covered stellar positions are all fainter than the isochrones, since starspots affect stellar luminosities more than colors.  The modeled curves demonstrate that the variation on $V$-band magnitude is more sensitive to $T_{\rm spot}/T_{\rm phot}$ (0.6 to 0.85) than $f_{\rm spot}$ (20\% to 60\%) at the warmest end of the isochrone. However, towards the cooler end, the brightness variation is controlled by $f_{\rm spot}$.

\begin{table}[!b]
\centering
\caption{Spot parameters}
\begin{tabular}{ccccc}
\hline
\hline
Group & $f_{\rm spot}$ & $T_{\rm spot} / T_{\rm phot}$ & $\Delta \log(L/\rm L_\odot)$ & $\Delta\log(L/\rm L_\odot)$ \\
  &   & & $T_{\rm phot} > 3800$ K & $T_{\rm phot} < 3800$ K \\

\hline
1 & 60\% & 0.6 & -0.35 & -0.36\\
2 & 60\% & 0.85 & -0.05 & -0.18 \\
3 & 40\% & 0.6 & -0.19 &  -0.20\\
4 & 40\% & 0.85 & -0.03 & -0.11 \\
5  & 20\% & 0.6 & -0.09 & -0.09 \\
6  & 20\% & 0.85 & -0.02 & -0.05 \\

 \hline
 \hline
 \label{tab:par-group}
\end{tabular}
\end{table}

A color-luminosity diagram is shown at the right panel of Figure~\ref{fig:cmd}, in which the stellar luminosities are calculated based on $V$-band magnitudes and bolometric corrections from \citet{Pecaut2013} based on $(B - V)$ colors. The absolute ages measured from H-R diagrams are affected by spots. First, evolutionary models that include spots will inflate radii of stars and raise the observed luminosity, relative to unspotted stars \citep{Jackson2014, Somers2015}. 
However, counter to this effect, when optical photometry or spectra of spotted stars are interpreted as emission from photospheres with a single temperature, then the measured temperature is hotter than the effective temperature and the luminosity is fainter than the real luminosity\footnote{For near-IR emission, the luminosity difference is suppressed but the temperature difference may still be significant.} In this way, the measured position of a spotted star should appear older than their age and more massive if the star is fit with multiple components.

\section{The contribution of spots to luminosity spreads of young clusters}

The duration of star formation in a single cluster should lead to an age and therefore a luminosity spread on H-R diagrams.  While luminosity spreads are measured for all young clusters \citep[e.g.][]{DaRio2010, Jose2017, Beccari2017}, the interpretation of the luminosity spread as an age spread is degenerate with any differences in evolution, stellar variability, and observational errors. The differences in evolution may be caused by a distribution of accretion histories \citep[e.g.][]{Hartmann1997, Hosokawa2011, Baraffe2017}, which should be minimal by the age of Pleiades, or by differences in interior structure caused by spots and by magnetic fields inhibiting convections \citep{Somers2015, Feiden2016}. In active star-forming regions, variability in accretion and extinction can also result as luminosity variation or spread \citep[e.g.][]{Venuti2015, Rodriguez2016, Contreras2017, Guo2018}.  However, the observational errors of the Pleiades should be minimal, since the Pleiades has no differential extinction, an accurate accounting of binarity \citep{Stauffer2007}, robust membership \citep{Lodieu2012}, and accurate distances from Gaia DR2 astrometry \citep{Gaia2018dr2catalog}.  The range of spot properties is likely the most significant uncertainty in both the observational measurements of stellar properties in the Pleiades and in the evolutionary models for stars of Pleiades age.

In this section, we first measure the empirical luminosity spread of the Pleiades. We then compare our spot simulations to the empirical luminosity spread to constrain the range in spot properties of low-mass stars in the Pleiades. Finally, we discuss challenges in applying these results to younger regions.

\begin{figure*}[!t]
\centering
\includegraphics[width=6.9in, angle=0]{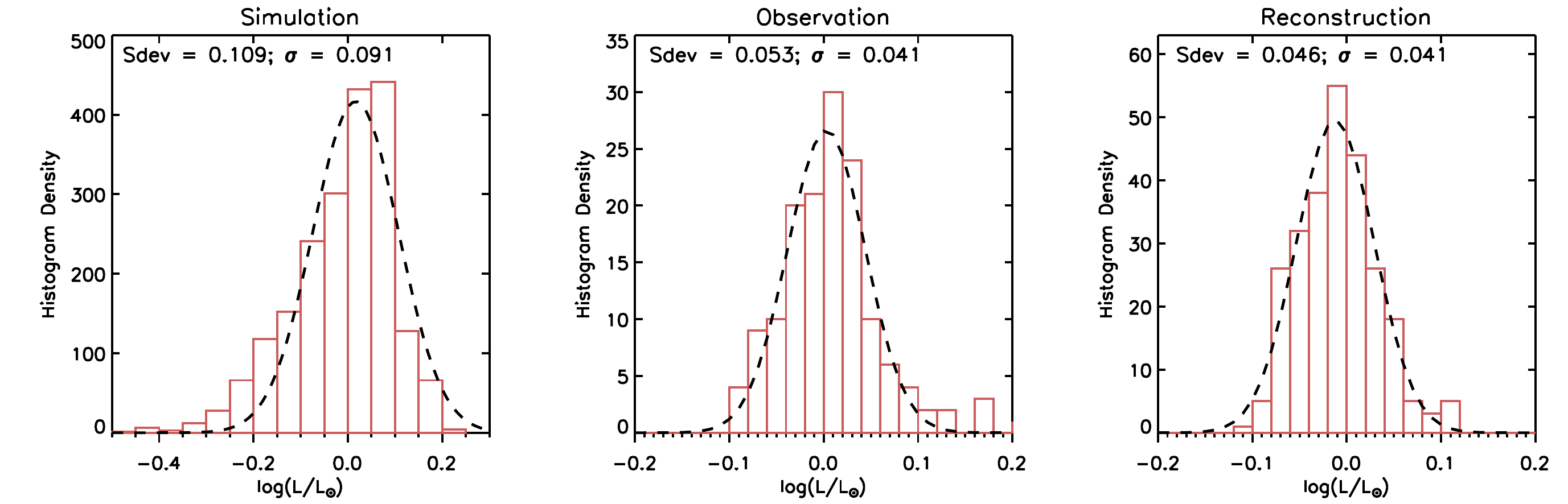}
\caption{ Histograms of luminosity spread calculated through H-R diagrams with fitted Gaussian profiles. Standard deviations of the luminosity spread as well as the $\sigma$ values from the Gaussian fittings are presented in each panel. {\it Left}: the luminosity spread is generated by simulated $L$ versus $(V - I)$. {\it Middle}:   observed luminosity spread of the Pleiades members \citep{Kamai2014} calculated based on $V$-band magnitude, individual distances from Gaia parallax, and bolometric correction based on $(V - I)$ color. {\it Right:} A special simulation to reconstruct the observed luminosity spread. The corresponding spot coverage is around 40\% under a normal distribution and the spot temperature is fixed to 60\% of the photospheric temperature.
}
\label{fig:lspread}
\end{figure*}

\subsection{The empirical luminosity spread of the Pleiades}

The Pleiades cluster includes over 2000 members \citep{Stauffer2007,Lodieu2012,Bouy2015}. In this work, we use the $B$, $V$ and $I_c$-band photometry obtained by \citet{Kamai2014}.  Of the 383 observed members, 61 are identified as binary or multiple systems from their location on the color-magnitude diagram and are ignored here.

We revise the membership classification of the stellar sample in \citet{Kamai2014} using the recent Gaia DR2 parallaxes and proper motions \citep{Gaia2016a, Gaia2018b, Luri2018}. By cross-matching the observed samples in \citet{Kamai2014} and Gaia DR2 database by on-sky stellar position, 311 single stars are detected with Gaia parallaxes. The cross-matched targets have parallaxes consistent with a Gaussian distribution at $7.3\pm0.2$ mas, or $137\pm4$ pc in distance, in which the errors are $1\sigma$ of the real depths including measurements uncertainties. 

 To avoid background and foreground contamination on the luminosity spread, we exclude the stars located outside 12 pc ($3\sigma$) from the median distance of the cluster of 137 pc.  An additional proper motion selection excludes 2 stars with proper motion 10 mas/yr $(5\sigma)$ away from the median values of 20 and -45 mas/yr.  In the end,  234 out of 311 single stars from \citet{Kamai2014} are identified as possible cluster members. The contamination rate from the non-members of the \citet{Kamai2014} members is at least $\sim 25$\%.

The selected 234 single Pleiades members are located in a color range of $0.4 < (V - I) < 3.3$, or earlier than M5V type\footnote{A large spread of stellar positions on the $V$ vs. $B - V$ diagrams is seen at the lower mass end \citep{Kamai2014}, which might be generated by chromospheric activities and the weak correlation between the color and stellar brightness at very low mass end. A smaller spread is also found on $(V - I)$. To avoid this bias, we set a lower mass limit at $(V - I) = 2.4$ (M3V), when calculating the luminosity spread.}. The absolute magnitude is calculated from the individual Gaia DR2 parallax distances for each star. The luminosity spread introduced by treating all members to a uniform 137 pc distance is 0.06 dex, according to the intrinsic spread of $\sigma = 4$ pc. While the luminosity spread based on the parallax uncertainties for each star is 0.02 dex.  The stellar luminosity is then calculated from the $V$-band absolute magnitude with a bolometric correction from the $(V - I)$ color. Then, an empirical isochrone is fit on the $\log(L/{\rm L_\odot}) - (V - I)$ space. The luminosity spread is calculated as the distance on the H-R diagram between each observed sample to the empirical isochrone on the luminosity axis. The distribution of observed luminosities, shown in the middle panel of Figure \ref{fig:lspread}, is well fit with a Gaussian profile with $\sigma = 0.05$ in the $\log(L/\rm L_\odot)$ space.  

\subsection{Constraining the range of spot properties of Pleiades stars and younger clusters}

To find out a possible origination of the observed luminosity spread, we compare two simulated luminosity spreads to observed samples in Figure~\ref{fig:lspread}. The first simulated cluster contains 2000 samples with spot coverages on each star evenly distributed between 1--40\%, as described in \S3.1. The distribution of simulated luminosities (the left panel of Figure~\ref{fig:lspread}) has a standard deviation, 0.1 dex, two times larger than the observed spread. The other simulated cluster marked as reconstruction in Figure~\ref{fig:lspread} contains 234 samples with spots covering an average of 40\% of the stellar surface in a Gaussian distribution with a standard deviation of 10\%. The spot temperatures are fixed to $0.6 T_{\rm phot}$. The ``reconstruction'' cluster yields a distribution of luminosities that is similar to the Pleiades. From this test, we learned that the luminosity distribution is controlled by the standard deviation in spot coverage. While the average spot coverage is not constrained. The low scatter in the stellar luminosity suggests that for the Pleiades, the spot coverage is similar for all members of a given stellar mass.

The standard deviation of 0.05 dex in luminosity at any single age is likely the minimum luminosity spread for young clusters.  The contribution of spots at younger ages may be larger. The variation amplitudes of the lightcurve of some weak-lined T Tauri stars can reach 0.3--0.5 mag, while others have little variation \citep[e.g.][]{Herbst1994, Grankin2008}. These large variations may be explained if younger stars have strong dipole magnetic fields \citep{Gregory2012} and therefore larger dispersions in estimated luminosities. Such large spots could help to explain the 0.2--0.3 dex luminosity spread seen in all young clusters  \citep[e.g.][]{DaRio2010,Preibisch2012,Herczeg2015}. By the age of the Pleiades, the magnetic structures for all stars of a given mass may have converged to similar spot coverages, leading to a smaller spread in luminosities.

\section{Summary}

A large observational discrepancy is seen on spot coverage between different methodologies. The spot coverage obtained from lightcurve amplitude is strongly underestimated. In this paper, we presented a series of Monte-Carlo simulations for spot distributions on 125-Myr-old young cluster members with masses ranging between 0.08 to 1.32 $\rm M_\odot$, including morphologies such as single, multiple, circumpolar and small spots.

In a simple one-star simulation, we learned that color and luminosity changes are dominated by spot filling factor. For multi-band lightcurves, $B$ and $V$-band are more sensitive to spots than $I$ and $K_s$-bands and the spot filling-factor is crucial to determine the variation amplitude. The color variations reach the maximum when $T_{\rm spot}/T_{\rm phot} \sim 0.85$.

For ``multiple spots'' and ``small spots'' configurations, a detection ratio ($\zeta = \Delta F_{\rm LC}/\Delta F_{\rm spot}$) is defined to quantify how much spot coverage is seen from the lightcurve. In the ``multiple spots'' morphology, half of the spot variations are measured on lightcurves. When the spot is distributed symmetrically along the longitude, the ``small spots'' case, 90\% of the spot behavior is hidden. The detection ratio between K2 lightcurve is 18\% and  LAMOST spectra, close to ``small spots'', suggesting most of the spot coverages are not detected.  

The spot-modified stellar loci are shown on color-color, color-magnitude, and color-luminosity (H-R) diagrams. In general, the spot-covered stars are redder and fainter than the photosphere. The luminosity spread of the observed samples is well reconstructed by cool spots on star surfaces with 10\% standard deviation in spot coverage. Large spots introduce luminosity spreads up to 0.1 dex on young clusters, while the measurements on spot coverage through K2 lightcurves are biased by their symmetrical distribution. 

{  Several effects, including faculae associated with dark spots, the inflation of stellar radii in result of the energy conservation, the latitudinal uneven distribution of star spots, the lifetime of spots, as well as other complicated configurations beyond and between our default settings, are not included in this work. However, our Monte-Carlo simulations provide a view of spot distribution, temperature, coverage and the corresponding luminosity spreads that well match the observational results.}

\section*{Acknowledgements}

We thank the anonymous referee for helpful comments and suggestions that improved the clarity and robustness of the results.
We thank Dr. Xiang-song Fang, Bharat Kumar, and the SAGE team leading by Prof. Gang Zhao in National Astronomical Observatory of China for proving their observational results on the LAMOST spectra. We thank Garrett Somers for valuable discussions.  ZG especially acknowledge the organizers and participants of the K2 Dwarf Stars and Clusters Workshop in January 2018 at the Boston University for helpful discussions related to ideas in this paper.

ZG and GJH are supported by general grants 11473005 and 11773002 awarded by the National Science Foundation of China.  This work has made use of data from the European Space Agency (ESA) mission {\it Gaia} (\url{https://www.cosmos.esa.int/gaia}), processed by the {\it Gaia} Data Processing and Analysis Consortium (DPAC,
\url{https://www.cosmos.esa.int/web/gaia/dpac/consortium}). Funding for the DPAC has been provided by national institutions, in particular, the institutions participating in the {\it Gaia} Multilateral Agreement.

\bibliographystyle{apj}
\bibliography{reference}
\end{document}